\documentclass[review]{elsarticle}
\usepackage[colorlinks, linkcolor=blue, anchorcolor=red, citecolor=green]{hyperref}
\usepackage{lineno,hyperref}
\modulolinenumbers[5]
\journal{XXX}

\newcommand{\beq}{\begin{equation}}
\newcommand{\eeq}{\end{equation}}
\newcommand{\bsubeq}{\begin{subequations}}
\newcommand{\esubeq}{\end{subequations}}
\newcommand{\beqn}{\begin{eqnarray}}
\newcommand{\eeqn}{\end{eqnarray}}
\newcommand{\fr}{\frac}
\newcommand{\lb}{\label}
\newcommand{\er}{\eqref}
\newcommand{\al}{\alpha}

\newcommand{\ka}{\kappa}

\newcommand{\om}{\omega}

\newcommand{\ga}{\gamma}

\newcommand{\ep}{\epsilon}
\newcommand{\la}{\lambda}

\newcommand{\bA}{\mathbf  A}
\newcommand{\bB}{\mathbf  B}
\newcommand{\bC}{\mathbf  C}

\newcommand{\bL}{\mathbf  L}

\newcommand{\bR}{\mathbf  R}

\newcommand{\pF}{\textbf{\emph{F}}}
\newcommand{\pG}{\textbf{\emph{G}}}
\newcommand{\pH}{\textbf{\emph{H}}}

\newcommand{\pP}{\textbf{\emph{P}}}

\newcommand{\pU}{\textbf{\emph{U}}}

\newcommand{\pf}{\textbf{\emph{f}}}

\newcommand{\pl}{\textbf{\emph{l}}}

\newcommand{\pr}{\textbf{\emph{r}}}
\newcommand{\pu}{\textbf{\emph{u}}}

\newcommand{\pat}{\partial}
\newcommand{\na}{\nabla}

\usepackage{amsmath}
\usepackage{multirow}
\usepackage{amsfonts,amssymb}
\usepackage{subfigure}

\begin{document}

\begin{frontmatter}

\title{\textbf{An extended hybrid numerical simulation of isotropic compressible turbulence}}

\author[mymainaddress]{L. Q. Liu}
\author[mysecondaryaddress]{J. C. Wang}
\author[mymainaddress]{Y. P. Shi}
\author[mymainaddress,mysecondaryaddress]{S. Y. Chen}
\author[mymainaddress,mythirdaryaddress]{X. T. He\corref{mycorrespondingauthor}}
\cortext[mycorrespondingauthor]{Corresponding author: Center for Applied Physics and Technology, Peking University, Beijing 100871, China. Tel.: +86-10-6275-3944; fax: +86-10-6275-3944;}
\ead{xthe@iapcm.ac.cn}

\address[mymainaddress]{State Key Laboratory for Turbulence and Complex Systems, Center for Applied Physics and Technology, College of Engineering, Peking University, Beijing 100871, China}
\address[mysecondaryaddress]{Department of Mechanics and Aerospace Engineering, Southern University of Science and Technology, Shenzhen 518055, China}
\address[mythirdaryaddress]{Institute of Applied Physics and Computational Mathematics, Beijing 100088, China}

\begin{abstract}
  This paper presents an extension of the hybrid scheme proposed by Wang et al. (\textit{J. Comput. Phys.} \textbf{229} (2010) 169-180) for numerical simulation of compressible isotropic turbulence to flows with higher turbulent Mach numbers. The scheme still utilizes an 8th-order compact scheme with built-in hyperviscosity for smooth regions and a 7th-order WENO scheme for highly compressive regions, but now both in their conservation formulations and for the latter with the Roe type characteristic-wise reconstruction. To enhance the robustness of the WENO scheme without compromising its high-resolution and accuracy, the recursive-order-reduction procedure is adopted, where a new type of reconstruction-failure-detection criterion is constructed. To capture the upwind direction properly in extreme conditions, the global Lax-Friedrichs numerical flux is used. In addition, a new form of cooling function is proposed, which is proved to be positivity-preserving. With these techniques, the new scheme not only inherits the good properties of the original one but also extends largely the computable range of turbulent Mach number, which has been further confirmed by numerical results.
\end{abstract}


\end{frontmatter}

\section{Introduction}\label{sec.1}

Compressible turbulence is of fundamental importance to a number of natural phenomena and industrial applications, including
interstellar medium \citep{Elmegreen2004, Scalo2004}, solar winds \citep{Alexandrova2013}, star-forming clouds in galaxies \citep{Mckee2007}, high-temperature reactive flows \citep{Pope1991}, supersonic aircraft design \citep{Ingenito2010} and inertial confinement fusion \citep{Lindl1998, He2016}. With increasing computational resources, direct numerical simulations of incompressible turbulent flows have been routinely conducted for many canonical boundary conditions and geometries. Similar developments for compressible flows are desired in order to provide parameterizations needed for modeling complex compressible turbulence in relevant applications.

While the pseudo-spectral method for incompressible isotropic homogeneous turbulence in a periodic domain has been well established \citep{Orszag1972}, such a standard method is no longer suitable for compressible turbulence at high Mach numbers due to the notorious Gibbs phenomenon \citep{Hewitt1979}. To overcome this barrier, two major strategies can be taken. The first one is the shock-fitting approach, which treats shock waves as genuine discontinuities, with dynamics governed by their own algebraic equations, and uses the Rankine-Hugoniot relations as a set of nonlinear boundary conditions to relate the states on the two sides of the discontinuity \citep{Moretti1987}. The second one is the shock-capturing approach, which uses the same discretization scheme at all points and achieves regularization through the addition of numerical dissipation. Although the former guarantees more accurate representations of shocked flows, it is merely feasible in cases where the shock topology is extremely simple and no shock wave forms during the calculation. Since our ultimate goal is to simulate compressible turbulence where shocklets form randomly, we discuss only the latter within the context of finite difference method (FDM) in this paper. Specifically, there are four major methods of this approach, namely, the classical shock-capturing methods \citep{Godunov1959, Leer1979, Colella1984, Toro2009, Courant1952, Harten1983, Harten1987, Liu1994}, the artificial viscosity methods \citep{Neumann1950, Jameson1981, Tadmor1989, Cook2004, Cook2005}, the nonlinear filtering methods \citep{Harten1978, Yee1999, Garnier2001, Yee2007}, and the hybrid methods \citep{Lee1997, Adams1996, Pirozzoli2002, Ren2003, Wang2010}. For a review on these methods, the paper of \citet{Pirozzoli2011} is highly recommended. At present, the hybrid methods are still the most superior one \citep{Johnsen2010} and thus are the main concern of this paper.

Briefly speaking, the hybrid methods are based on the idea of endowing a baseline spectral-like scheme with shock-capturing capability through local replacement with a classical shock-capturing scheme, where the shock sensor plays a key role. Along this direction, some progresses have been made in the past two decades. In particular, \citet{Adams1996} first considered a truly adaptive hybrid discretization, consisting of a baseline 5th-order compact upwind (CU) scheme coupled with a 5th-order essentially non-oscillatory (ENO) scheme, where the shock sensor is based on the local gradient of the flux vector components. \citet{Pirozzoli2002} expanded this method by transforming it into a fully conservative formulation, replacing the ENO with weighted essentially non-oscillatory (WENO), and using the local density gradient as the shock sensor. This method was further improved by \citet{Ren2003}, who used the Roe type characteristic-wise reconstruction and introduced a complex weight function to gradually switch between CU and WENO at the interface. \citet{Zhou2007} introduced a new family of CU and combined this with WENO. There are also several other studies that combine the usual, non-compact scheme and WENO \citep{Hill2004, Kim2005, Larsson2007}, but they have the very similar issues in shock detection and the interface treatments as discussed above.

Recently, \citet{Wang2010} developed a novel hybrid numerical scheme with built-in hyperviscosity that is applicable to the numerical simulation of compressible isotropic turbulence (CIT) with relatively high turbulent Mach number $M_t \lesssim 1.0$. This scheme utilizes a 7th-order WENO scheme for highly compressive regions and an 8th-order compact central (CC) scheme for smooth regions, with the shock sensor being the shocklet detection algorithm given by \citet{Samtaney2001}. In addition, a numerical hyperviscosity formulation is proposed to remove the alias error without compromising numerical accuracy. With this scheme (hereafter we call it Wang's scheme for short), they made a thorough study of CIT at $M_t \lesssim 1.0$, including but not limited to effect of shocklets on the velocity gradients \citep{Wang2011}, effects of local compressibility on the statistical properties and structures of velocity gradients \cite{Wang2012-jfm}, scaling and statistics of velocity structure functions \citep{Wang2012-prl}, and shocklets-particle interaction \citep{Yang2014}, to name a few. For recent progress in this direction, please see the review of \citet{Chen2015}.

Needless to say, these progresses have deepened our understandings of compressible turbulence. But their computable range of $M_t$ is still too narrow. As a result, the thermodynamic process has not been fully activated since the kinetic energy is still much smaller than the internal energy. In fact, if the averaged kinetic energy equals to the averaged internal energy, there must be $M_t \approx 1.9$ for polytropic gas with specific heat ratio $\ga = 1.4$. However, for such a large $M_t$, some numerical instability issues may bring up, making Wang's scheme blow-up. On the other hand, although the available $M_t$ in literature has been as large as 17 for isothermal Euler turbulence \citep{Federrath2013}, it is scarcely larger than unity for the simulated turbulence of viscous and compressible fluid. For example, the largest $M_t$ for this type of flow in literature is 0.6 for compact scheme \citep{Jagannathan2016}, 0.8 for optimized WENO scheme \citep{Pirozzoli2004}, 0.885 for gas kinetic method \citep{Lee2009}, $1.02$ for localized artificial method \citep{Xia2016}, and $1.03$ for hybrid approach \citep{Wang2012-prl}. Therefore, our present aim is to extend Wang's scheme to a wider range of $M_t$.

The paper is organized as follows. Some backgrounds will be first presented in \S~\ref{sec.2}. Then, by improving its robustness and proposing a new form of cooling function, Wang's scheme will be extended to flows with a higher $M_t$ in \S~\ref{sec.3}, which inherits the good properties of the former. A series of numerical tests are then shown in \S~\ref{sec.4} to illustrate the improvement and extension of the new schemes. Finally, main conclusions are summarized in \S~\ref{sec.5}.

\section{Backgrounds}\label{sec.2}

\subsection{Governing equations for compressible turbulent flow}\label{sec.2.1}

In this study we consider a compressible turbulence of a perfect gas driven and maintained by large-scale momentum forcing $\pf$ and thermal forcing $\Lambda$. The governing equations are
\begin{equation}\label{NS-3d}
  \fr{\pat \pU}{\pat t} + \fr{\pat \pF}{\pat x} + \fr{\pat \pG}{\pat y} + \fr{\pat \pH}{\pat z} = \pP + \fr{\pat \pF_v}{\pat x} + \fr{\pat \pG_v}{\pat y} + \fr{\pat \pH_v}{\pat z},
\end{equation}
where
\begin{equation}
  \pU = \left( \begin{array}{c} \rho \\ \rho u \\ \rho v \\ \rho w \\ E \\ \end{array} \right), \quad
  \pP = \left( \begin{array}{c} 0 \\ f_x \\ f_y \\ f_z \\
  \Lambda + \pf\cdot\pu \\ \end{array} \right)
\end{equation}
are the conservative variable and the source term, respectively,
\begin{equation}
  \pF = \left( \begin{array}{c} \rho u \\ \rho u^2 + p \\ \rho uv \\ \rho uw \\ u (E + p) \\ \end{array} \right), \quad
  \pG = \left( \begin{array}{c} \rho v \\ \rho v u \\ \rho v^2 + p \\ \rho v w \\ v (E + p) \\ \end{array} \right), \quad
  \pH = \left( \begin{array}{c} \rho w \\ \rho wu \\ \rho wv \\ \rho w^2 + p \\ w (E + p) \\ \end{array} \right)
\end{equation}
are the inviscid fluxes along $x$, $y$, and $z$ directions, respectively, and
\begin{equation}
  \pF_v = \left( \begin{array}{c} 0 \\ \tau_{xx} \\ \tau_{xy} \\ \tau_{xz} \\ \tau_{xi} u_i +  \ka \fr{\pat T}{\pat x}\\ \end{array} \right), \quad
  \pG_v = \left( \begin{array}{c} 0 \\ \tau_{yx} \\ \tau_{yy} \\ \tau_{yz} \\ \tau_{yi} u_i +  \ka \fr{\pat T}{\pat y}\\ \end{array} \right), \quad
  \pH_v = \left( \begin{array}{c} 0 \\ \tau_{zx} \\ \tau_{zy} \\ \tau_{zz} \\ \tau_{zi} u_i +  \ka \fr{\pat T}{\pat z}\\ \end{array} \right)
\end{equation}
are the viscous fluxes along $x$, $y$, and $z$ directions, respectively. In particular, under the Stokes assumption, the viscous stress tensor $\{ \tau_{ij} \}$ is
\begin{equation}
  \{ \tau_{ij} \} = \mu  \left(
  \begin{split}
   & 2 \fr{\pat u}{\pat x} - \fr{2}{3} \theta & & \fr{\pat u}{\pat y} + \fr{\pat v}{\pat x} & & \fr{\pat u}{\pat z} + \fr{\pat w}{\pat x} \\
  & \fr{\pat u}{\pat y} + \fr{\pat v}{\pat x} & & 2 \fr{\pat v}{\pat y} - \fr{2}{3} \theta & & \fr{\pat v}{\pat z} + \fr{\pat w}{\pat y} \\
  & \fr{\pat u}{\pat z} + \fr{\pat w}{\pat x} & & \fr{\pat v}{\pat z} + \fr{\pat w}{\pat y} & & 2 \fr{\pat w}{\pat z} - \fr{2}{3} \theta \\
  \end{split}
  \right),
\end{equation}
where $\theta = \na \cdot \pu$ is the dilatation and $\mu$ is the dynamic viscosity.

For the perfect gas, the viscosity $\mu$, as well as the heat transfer coefficient $\ka$, increases with temperature $T$, which can be well modeled by Sutherland's law \citep{Sutherland1893},
\begin{equation}\label{8}
  \fr{\mu}{\mu_0} = \fr{\ka}{\ka_0} = \fr{1 + 0.4042}{\displaystyle \fr{T}{T_0} +0.4042} \left( \fr{T}{T_0} \right)^{1.5},
\end{equation}
where subscript 0 denotes reference value. For air, $\mu_0 = 1.716 \times 10^{-5} \rm kg/ m \cdot s$, $T_0 = 273.15 \rm K$, and \er{8} is valid when $0.37 \le T/T_0 \le 7.0$; while $\ka_0$ can be obtained from the definition $Pr = \mu_0 c_p/\ka_0$, where $Pr$ is the Prandtl number that can be approximated as 0.7 when $0.73 \le T/T_0 \le 9.2$, and $c_p$ is the specific heat at constant pressure, which, for perfect gas, also depends only on $T$. For moderate large $M_t$, the local temperature can be very small due to the energy exchange between kinetic energy and internal energy. For instance, when $M_t = 2.3$, the minimum value of $T/T_0$ can be as small as 0.13, which may violate the Sutherland law. On the other hand, for extremely large $M_t \ge 5$, the assumption of perfect gas may be invalid and the real gas effect should be taken into account \citep{Lunev2009}. These topics are of course very important, however, they are out of the scope of this paper.

Similar to forcing incompressible turbulence \citep{Chen1993}, the momentum forcing field $\pf$ is constructed in the Fourier space by fixing the kinetic energy $E(k)$ per unit mass within the two lowest wave number shells, $0.5 < k \le 1.5$ and $1.5 < k \le 2.5$, to prescribed values consistent with the $k^{-5/3}$ kinetic energy spectrum. Other ways to add the forcing term can be found in literatures, for example, \citet{Jagannathan2016}. Since momentum forcing increases the total energy of the system, one has to devise a mechanism to remove energy to maintain a stationary state. This can be done by removing energy through the cooling function $\Lambda$ either such that the mean internal energy is kept constant or homogeneously and with a value equal to the average energy input at every step \citep{Jagannathan2016}. In particular, following \citet{Passot1995}, \citet{Wang2010} proposed the following cooling function
\begin{equation}\label{38}
  \Lambda = a T^b,
\end{equation}
where $a$ and $b$ are two parameters. Three different values of $b$, namely, $b = 0, 2, 4$, were considered by \citet{Wang2010}, and no significant differences were found at their investigated range of $M_t$. When $M_t$ is larger than unity, however, these cooling functions will lead to negative temperature. Thus, a new form of cooling function is desiring, which is to be done in \S~\ref{sec.cooling} below.

\subsection{The 7th-order WENO scheme}\label{sec.2.2}

For completeness the 7th-order WENO scheme of \citet{Balsara2000} is listed here. Consider the scalar hyperbolic conservation law given by
\begin{equation}\label{Ren-03-1}
  \fr{\pat u}{\pat t} + \fr{\pat f(u)}{\pat x} = 0
\end{equation}
with proper initial and boundary conditions. By the hyperbolicity of \er{Ren-03-1}, $\pat f/\pat u$ is a real function of $u$. Let $\{I_j\}$ be a uniform partition of the solution domain in space, where $I_j = \{x_{j-\fr{1}{2}}, x_{j+\fr{1}{2}} \}$ and $x_{j+ \fr{1}{2}} - x_{j-\fr{1}{2}} = \Delta x$. The semi-discrete conservative finite difference scheme of \er{Ren-03-1} can be written as
\begin{equation}\label{Ren-03-2}
  \fr{\pat u}{\pat t} + \fr{\hat f_{j+\fr{1}{2}} - \hat f_{j-\fr{1}{2}}}{\Delta x} = 0,
\end{equation}
where $\hat f_{j+\fr{1}{2}}$ is the numerical flux function. Then, if
\begin{equation}\label{Ren-03-3}
  \fr{\hat f_{j+\fr{1}{2}} - \hat f_{j-\fr{1}{2}}}{\Delta x} = \left. \fr{\pat f}{\pat x} \right|_j + O(\Delta x^k),
\end{equation}
the scheme is called $k$-th order accurate in space.

In general, the flux $f$ can be split into a positive part and a negative part,
\begin{equation}\label{fmp}
  f = f^+ + f^-,
\end{equation}
which can be carried out either by the Boltzmann approach (e.g., Lax-Friedrichs flux, Steger-Warming flux) or the Riemann approach (e.g., Godunov flux, Roe flux). A brief discussion about the upwind capture method is given in \S~\ref{sec.upwind} below. In particular, for the positive flux, there is
\begin{equation}\label{2.10-96}
  \hat f^+_{j+1/2} = \sum_{k=0}^{3} \om_k q^+_k(x_{j+1/2}; f^+_{j+k-3}, \cdots, f^+_{j+k}),
\end{equation}
where $q^+_k(x_{j+1/2}; f^+_{j+k-3}, \cdots, f^+_{j+k})$ is obtained by a 3rd-order polynomial reconstruction of $f^+(u(x_{j+1/2}))$ on $k$th set of candidate stencils $S_k$,
\begin{equation}
 \left\{ \begin{split}
 & q^+_0(x_{j+1/2}) = \fr{1}{12} (-3 f^+_{j-3} + 13 f^+_{j-2} - 23 f^+_{j-1} + 25 f^+_{j}), \\
 & q^+_1(x_{j+1/2}) = \fr{1}{12} ( f^+_{j-2} - 5 f^+_{j-1} + 13 f^+_{j} + 3 f^+_{j+1}), \\
 & q^+_2(x_{j+1/2}) = \fr{1}{12} (-f^+_{j-1} + 7 f^+_{j} + 7 f^+_{j+1} - f^+_{j+2}), \\
 & q^+_3(x_{j+1/2}) = \fr{1}{12} ( 3 f^+_{j} + 13 f^+_{j+1} - 5 f^+_{j+2} + f^+_{j+3}),
 \end{split} \right.
\end{equation}
and $\om_k$ is the weight, which can be written as
\begin{equation}\label{2.15-96}
  \om_k = \fr{\al_k}{ \sum_{n=0}^3 \al_n}, \quad
  \al_k = \fr{C_k}{(\ep + \mathrm{IS}_k)^p}.
\end{equation}
Here, $C_k$ is the optimal weight,
\begin{equation}\label{c4}
  C_0 = \fr{1}{35}, \quad
  C_1 = \fr{12}{35}, \quad
  C_2 = \fr{18}{35}, \quad
  C_3 = \fr{4}{35},
\end{equation}
the power $p = 2$, $\ep = 10^{-6}$ is a positive real number which is introduced to avoid the denominator becoming zero, and $\mathrm{IS}_k$ is a smoothness measurement of the flux function on the $k$th candidate stencil,
\begin{equation}\label{3.1-96}
  \mathrm{IS}_k = \sum_{l=1}^{3} \int_{x_{j-1/2}}^{x_{j+1/2}} \Delta x^{2l-1} \left[ q_k^{(l)} \right]^2 \mathrm{d}x,
\end{equation}
i.e.,
\begin{equation}
 \mathrm{IS}_k = \left( q_k^{(1)}(x_j) \right)^2 + \fr{13}{12} \left( q_k^{(2)}(x_j) \right)^2 + \fr{1043}{960} \left( q_k^{(3)}(x_j) \right)^2 + \fr{1}{12} q_k^{(1)}(x_j) q_k^{(3)}(x_j),
\end{equation}
where $k=0, 1, 2, 3$ and
\begin{equation}
 \left\{ \begin{split}
 & q_0^{(1)}(x_j) = \fr{1}{6} (-2 f_{j-3} + 9 f_{j-2} - 18 f_{j-1} + 11 f_{j}), \\
 & q_1^{(1)}(x_j) = \fr{1}{6} ( f_{j-2} - 6 f_{j-1} + 3 f_{j} + 2 f_{j+1}), \\
 & q_2^{(1)}(x_j) = \fr{1}{6} (-2 f_{j-1} - 3 f_{j} + 6 f_{j+1} -  f_{j+2}),  \\
 & q_3^{(1)}(x_j) = \fr{1}{6} (-11 f_{j} + 18 f_{j+1} - 9 f_{j+2} + 2 f_{j+3}),
 \end{split} \right.
\end{equation}
\begin{equation}
 \left\{ \begin{split}
  & q_0^{(2)}(x_j) = -f_{j-3} + 4 f_{j-2} - 5 f_{j-1} + 2 f_{j}, \\
  & q_1^{(2)}(x_j) = f_{j-1} - 2f_{j} + f_{j+1}, \\
  & q_2^{(2)}(x_j) = f_{j} - 2f_{j+1} + f_{j+2}, \\
  & q_3^{(2)}(x_j) = 2f_{j} -5 f_{j+1} +4 f_{j+2} - f_{j+3},
 \end{split} \right.
\end{equation}
\begin{equation}
 \left\{ \begin{split}
  & q_0^{(3)}(x_j) = -f_{j-3} + 3 f_{j-2} - 3 f_{j-1} + f_{j}, \\
  & q_1^{(3)}(x_j) = - f_{j-2} + 3 f_{j-1} - 3 f_{j} + f_{j+1}, \\
  & q_2^{(3)}(x_j) = - f_{j-1} + 3 f_{j} - 3 f_{j+1} + f_{j+2},\\
  & q_3^{(3)}(x_j) = -f_{j} + 3 f_{j+1} - 3 f_{j+2} + f_{j+3}.
 \end{split} \right.
\end{equation}

Note that the above arguments are only suitable for positive flux $f^+_{j+1/2}$. Nevertheless, the negative flux $f^-_{j+1/2}$ is the same as $f^+_{j+1/2}$ but with the corresponding $j+k$ replaced by $j+1-k$.

\subsection{The hybrid scheme of Wang et al. (2010)}

\citet{Wang2010} combined the above 7th-order WENO scheme \citep{Balsara2000} for the shock regions and an 8th-order CC scheme \citep{Lele1992} for smooth regions to treat the advection terms in the governing equations, where the shock front is identified by spatial points with highly negative local dilatation as defined by $\theta / \theta_{\rm rms} < -3$ \citep{Samtaney2001}. {\color{black} Here $\theta_{\rm rms}$ denotes the root-mean-square of the dilatation field.} Additional six grid points are added on both left and right in each spatial direction immediately outside the front to depress the potential oscillation due to the interface. Thus, they obtained
\begin{equation}\label{29-W}
  \fr{3}{8} \pF'_{j-1} + \pF'_{j} + \fr{3}{8} \pF'_{j + 1} = \fr{ \pF_{j+1/2}^{\rm Hybrid} - \pF_{j-1/2}^{\rm Hybrid} }{\Delta x},
\end{equation}
where $\pF'$ is the $x$-derivative of the physical flux,
\begin{equation}\label{30-W}
  \pF_{j+1/2}^{\rm Hybrid} \equiv \left\{
  \begin{split}
   & \pF_{j+1/2}^{\rm Compact} & &\quad \textrm{for smooth regions,} \\
   & \pF_{j+1/2}^{\rm WENO}  & &\quad \textrm{for shock regions,}  \\
   & \fr{1}{2} \left( \pF_{j+1/2}^{\rm Compact} + \pF_{j+1/2}^{\rm WENO} \right) & &\quad \textrm{at the joint,}
  \end{split}
  \right.
\end{equation}
and
\begin{equation}
 \left\{ \begin{split}
  & \pF_{j+1/2}^{\rm Compact} = \fr{398}{480} (\pF_{j} + \pF_{j+1}) + \fr{23}{480} (\pF_{j-1} + \pF_{j+2}) -\fr{1}{480} (\pF_{j-2} + \pF_{j+3}), \\
  & \pF_{j+1/2}^{\rm WENO} = \fr{3}{8} \hat \pF^{\rm WENO}_{j-\fr{1}{2}} + \hat \pF^{\rm WENO}_{j+ \fr{1}{2}} + \fr{3}{8} \hat \pF^{\rm WENO}_{j + \fr{3}{2}}.
 \end{split} \right.
\end{equation}

On the one hand, the upwind direction was captured by the Steger-Warming flux splitting method \citep{Steger1981}. The viscous term in the momentum equation was handled by a 6th-order non-compact central scheme. The same method was used for the viscous dissipation term in the energy equation. However, the thermal diffusion term in the energy equation was treated by an 8th-order compact scheme, since it carries several other parameters. The time marching was performed by an explicit low storage 2nd-order Runge-Kutta technique \citep{Shu1988}.

On the other hand, since the central scheme is not stable due to the existence of alias error \citep{Phillips1959}, \citet{Wang2010} proposed a natural numerical viscosity treatment, in line with compact schemes, that will dissipate the unwanted small-scale fluctuations without compromising the accuracy of the scheme. The novel aspect of the implementation is that the numerical viscosity term has a high-order spatial structure similar to the discretization error of the overall numerical method. This is accomplished by taking the difference of two compact algorithms of 2nd-order spatial derivatives. The first is based on a CC scheme applied to the 1st-order spatial derivatives, but applied twice to yield the 2nd-order derivatives, while the second is a CC scheme applied directly to the 2nd-order derivatives. This technique will also be adopted in the present paper and more details can be found in \citet{Wang2010} .

\section{The improved and extended hybrid scheme}\label{sec.3}

Before proceeding, we first remark that there are two possibilities that may degrade the high-resolution of the scheme of \citet{Wang2010}. One is that their scheme is not in truly conservation formulation as they asserted. This can be seen clearly from \er{29-W}, where the derivative of physical flux rather than numerical flux is solved. In fact, the scheme in truly conservation formulation is
\begin{equation}\label{29}
  \fr{3}{8} \hat \pF_{j-\fr{1}{2}} + \hat \pF_{j+ \fr{1}{2}} + \fr{3}{8} \hat \pF_{j + \fr{3}{2}} = \pF_{j+1/2}^{\rm Hybrid},
\end{equation}
where $\hat \pF$ is the numerical flux and $\pF_{j+1/2}^{\rm Hybrid}$ is still given by \er{30-W}. This can be obtained by following the tactic of \citet{Pirozzoli2002}, of which the concept can be dated back to \citet{Lax1954}. Our numerical results show that this improvement has a negligible effect on the statistical properties of the flow field. However, for the instantaneous distribution of flow field there are some significant differences when shocklets exist (figure not shown). For example, when $M_t = 0.7$ there may be a relative larger discontinuity of the result obtained by the original scheme, which may be spurious and could lead to spurious oscillations. Thus, we prefer \er{29} rather than \er{29-W} as our final formulation.

The other one is that the WENO scheme they adopted is based on the conservation-wise reconstruction rather than the characteristic-wise reconstruction, which may also lead to spurious oscillations. This can be improved by following the procedure of \citet{Ren2003}, of which the technique can be traced back to \citet{Harten1986}, where an implicit assumption, i.e., there is at least one smooth stencil, was used to prove the ENO property of the ENO scheme. This assumption, however, can never be satisfied in such problems where shock waves intersect with each other. To depress potential oscillations due to the violation of this assumption, \citet{Harten1987} first recommended to use this kind of reconstruction. Similar to the first case, this improvement has also a negligible effect on the statistical properties of the flow field for $M_t < 1$. For extreme conditions (say, $M_t \gg 1$) where all flow processes are activated and so are their interactions, the scheme incorporated with conservation-wise reconstruction may fail due to numerical instability. Thus, we choose the characteristic-wise reconstruction, of which the details are given in \S~\ref{sec.character} and some one-dimensional numerical results are given in \S~\ref{sec.4.1}.

Characteristic-wise reconstruction alone, however, can never totally depress numerical oscillations and avoid blow-up. This is more serious in high-order schemes. Since these oscillations are mainly due to the Gibbs phenomena, more robust WENO schemes are desiring. Realizing that the lower-order stencils are more likely to have at least one smooth stencil than that of a higher one, it is natural to take a lower-order scheme if there is no smooth stencil of the higher-order. This can be carried out recursively until the 1st-order scheme if there are discontinuities everywhere, which can be designed monotonic and of which the solution has been mathematically proved to exist at least for one-dimensional flow. This is the so-called recursive-order-reduction (ROR) method \citep{Titarev2004, Gerolymos2009}, where the key ingredient is the reconstruction-failure-detection (RFD) criterion. However, the available criteria are mainly based on the absolute finite difference of the density or pressure, which is somewhat empirical and may be problematic for complex applications. Thus, more natural criterion is very helpful. Along this direction, we observe that the idea of bound-preserving limiters can be used as our criterion \citep{Shu2016}. In particular, \citet{Hu2013} proposed an \textit{a posteriori} approach, which first detects the critical numerical fluxes that may lead to negative pressure or density and then imposes a positivity-preserving flux limiter to correct the fluxes. Although this approach is simple and has a strong mathematical foundation, it still can not be directly utilized since wrong solution may be obtained. Nevertheless, the essential idea of positivity-preserving can be exploited as the RFD criterion, which will be incorporated in the improved hybrid scheme in \S~\ref{sec.positive}.

During our numerical tests, we also found that, in the framework of flux vector splitting (FVS) approach, the upwind direction can not be properly captured if local type splitting is used when $M_t > 1$, such as the Steger-Warming (SW) flux splitting used by \citet{Wang2010} or local Lax-Friedrichs (LF) flux splitting, which will also lead the simulation to blow up even for the 1st-order upwind scheme. However, for global one, such as the global LF flux splitting, it works well. Thus, the global LF flux will be adopted in our scheme and a physical explanation is given in \S~\ref{sec.upwind}. Other types of global flux should also work well, which, however, will not be tested in the present paper.

In addition, as remarked in \S~\ref{sec.2.1}, the cooling function of \citet{Wang2010} may also bring up some numerical instability issues when $M_t>1$. That is, it may lead to negative temperature and thus makes the simulation blow up. This is indeed the case for the uniform cooling, where the simulation blows up even for the 1st-order monotonic schemes such as the Godunov scheme when $M_t = 2$. To overcome this difficulty, a positivity preserving cooling function is proposed in \S~\ref{sec.cooling}.

\subsection{Characteristic-wise reconstruction}\label{sec.character}

In the present paper, a hybrid compact-WENO scheme is proposed which couples the conservation-wise compact sub-scheme with the characteristic-wise WENO sub-scheme. The evaluation of the numerical flux functions for the characteristic-wise compact scheme has been proposed by \citet{Ren2003}. Since it is relatively new, we repeat it here for reference. For completeness, some useful equations are also listed here \citep{Toro2009}.

Define
\begin{equation}\label{H-k-gamma-3d}
  H = k + \fr{a^2}{\hat \ga}, \quad
  k = \fr{1}{2} (u^2 + v^2 + w^2), \quad
  a = \sqrt{ \fr{\ga p}{\rho} }, \quad
  \hat \ga = \ga -1.
\end{equation}
Then we have (see \er{NS-3d} for reference)
\begin{equation}\label{A-3d}
  \bA \equiv \fr{\pat \pF}{\pat \pU} =
  \left(
    \begin{array}{ccccc}
      0 & 1 & 0 & 0 & 0 \\
      \hat \ga k - u^2 & (3-\ga) u & -\hat \ga v & -\hat \ga w & \hat \ga \\
      -uv & v & u & 0 & 0 \\
      -uw & w & 0 & u & 0 \\
      \left(\hat \ga k- H \right) u & H - \hat \ga u^2 & -\hat \ga uv & -\hat \ga uw & \ga u \\
    \end{array}
  \right),
\end{equation}
\begin{equation}\label{B-3d}
  \bB \equiv \fr{\pat \pG}{\pat \pU} =
  \left(
    \begin{array}{ccccc}
      0 & 0 & 1 & 0 & 0 \\
      -vu & v & u & 0 & 0 \\
      \hat \ga k - v^2 & -\hat \ga u & (3-\ga) v & -\hat \ga w & \hat \ga \\
      -vw & 0 & w & v & 0 \\
      \left(\hat \ga k- H \right) v & -\hat \ga vu & H - \hat \ga v^2 & -\hat \ga vw & \ga v \\
    \end{array}
  \right),
\end{equation}
\begin{equation}\label{C-3d}
  \bC \equiv \fr{\pat \pH}{\pat \pU} =
  \left(
    \begin{array}{ccccc}
      0 & 0 & 0 & 1 & 0 \\
      -wu & w & 0 & u & 0 \\
      -wv & 0 & w & v & 0 \\
      \hat \ga k - w^2 & -\hat \ga u & -\hat \ga v & (3-\ga) w &\hat \ga \\
      \left(\hat \ga k- H \right) w & -\hat \ga wu & -\hat \ga wv & H - \hat \ga w^2 &\ga w \\
    \end{array}
  \right).
\end{equation}
The eigenvalues of the Jacobian matrices $\bA, \bB, \bC$ are
\begin{equation}\label{lambda-A}
\left. \begin{split}
  & \la^A_1 = u-a, & & \la^A_{2,3,4} = u, & & \la^A_5 = u+a, \\
  & \la^B_1 = v-a, & & \la^B_{2,3,4} = v, & & \la^B_5 = v+a, \\
  & \la^C_1 = w-a, & & \la^C_{2,3,4} = w, & & \la^C_5 = w+a, \\
\end{split} \right.
\end{equation}
and the corresponding right and left eigenvectors are
\begin{equation}\label{RA-3d}
   \bR_A = \left( \begin{array}{ccccc}
  1 & 1 & 0 & 0& 1 \\
  u-a & u & 0& 0& u+a \\
  v & v & 1 & 0 & v \\
  w & w & 0 & 1 & w\\
  H - ua & k & v & w & H + ua \\
  \end{array} \right),
\end{equation}
\begin{equation}\label{RB-3d}
   \bR_B = \left( \begin{array}{ccccc}
  1 & 0 & 1 & 0 & 1 \\
  u & 1 & u & 0 & u \\
  v-a & 0 & v & 0 & v+a \\
  w & 0 & w & 1 & w \\
  H - va & u & k & w & H + va \\
  \end{array} \right),
\end{equation}
\begin{equation}\label{RC-3d}
   \bR_C = \left( \begin{array}{ccccc}
  1 & 0 & 0 & 1 & 1 \\
  u & 1 & 0 & u & u \\
  v & 0 & 1 & v & v \\
  w-a & 0 & 0 & w & w+a \\
  H - wa & u & v & k & H + wa \\
  \end{array} \right),
\end{equation}
and
\begin{equation}\label{LA-3d}
   \bL_A = \fr{\hat \ga}{2 a^2} \left( \begin{array}{ccccc}
  k + \fr{a}{\hat \ga} u & -u -\fr{a}{\hat \ga} & -v & -w & 1 \\
  \fr{2a^2}{\hat \ga} - 2k & 2u & 2v & 2w & -2 \\
  -\fr{2a^2}{\hat \ga}v & 0 & \fr{2a^2}{\hat \ga} & 0 & 0 \\
  -\fr{2a^2}{\hat \ga}w & 0 & 0 & \fr{2a^2}{\hat \ga} & 0 \\
  k - \fr{a}{\hat \ga} u & -u + \fr{a}{\hat \ga} & -v & -w & 1 \\
  \end{array} \right),
\end{equation}
\begin{equation}\label{LB-3d}
   \bL_B = \fr{\hat \ga}{2 a^2} \left( \begin{array}{ccccc}
  k + \fr{a}{\hat \ga} v & -u & -v -\fr{a}{\hat \ga} & -w & 1 \\
  -\fr{2a^2}{\hat \ga}u & \fr{2a^2}{\hat \ga} & 0 & 0 & 0 \\
  \fr{2a^2}{\hat \ga} - 2k & 2u & 2v & 2w & -2 \\
  -\fr{2a^2}{\hat \ga}w & 0 & 0 & \fr{2a^2}{\hat \ga} & 0 \\
  k - \fr{a}{\hat \ga} v & -u & -v + \fr{a}{\hat \ga} & -w & 1 \\
  \end{array} \right),
\end{equation}
\begin{equation}\label{LC-3d}
   \bL_C = \fr{\hat \ga}{2 a^2} \left( \begin{array}{ccccc}
  k + \fr{a}{\hat \ga} w & -u & -v & -w -\fr{a}{\hat \ga} & 1 \\
  -\fr{2a^2}{\hat \ga}u & \fr{2a^2}{\hat \ga} & 0 & 0 & 0 \\
  -\fr{2a^2}{\hat \ga}v & 0 & \fr{2a^2}{\hat \ga} & 0 & 0 \\
  \fr{2a^2}{\hat \ga} - 2k & 2u & 2v & 2w & -2 \\
  k - \fr{a}{\hat \ga} w & -u & -v & -w + \fr{a}{\hat \ga} & 1 \\
  \end{array} \right),
\end{equation}
respectively.

On the other hand, the local characteristic decompositions are adopted. Take the flux along $x$-direction as an example, there is,
\begin{equation}\label{A-2}
  \bA_{j+1/2} = \left. \fr{\pat \pF(\pU)}{\pat \pU} \right|_{j+1/2},
\end{equation}
where $\pU_{j+1/2}$ is the Roe average \citep{Roe1981},
\begin{equation}\label{Roe-aver}
\left\{\begin{split}
  u_{j+1/2} =& \fr{ \sqrt{\rho_j} u_j + \sqrt{\rho_{j+1}} u_{j+1} }{ \sqrt{\rho_j} + \sqrt{\rho_{j+1}} }, \quad
  v_{j+1/2} =  \fr{ \sqrt{\rho_j} v_j + \sqrt{\rho_{j+1}} v_{j+1} }{ \sqrt{\rho_j} + \sqrt{\rho_{j+1}} }, \\
  w_{j+1/2} =& \fr{ \sqrt{\rho_j} w_j + \sqrt{\rho_{j+1}} w_{j+1} }{ \sqrt{\rho_j} + \sqrt{\rho_{j+1}} }, \quad
  H_{j+1/2} = \fr{ \sqrt{\rho_j} H_j + \sqrt{\rho_{j+1}} H_{j+1} }{ \sqrt{\rho_j} + \sqrt{\rho_{j+1}} }, \\
  a_{j+1/2} =& \sqrt{ (\ga-1) \left[ H_{j+1/2} - \fr{1}{2} \left(u_{j+1/2}^2 + v_{j+1/2}^2 + w_{j+1/2}^2 \right) \right]}.
\end{split}\right.
\end{equation}
Then, by projecting the fluxes onto the characteristic plane, there is
\begin{equation}\label{F-pm-c-2}
  F^\pm_{s; j+k} = \fr{1}{2} \pl_{s, j+1/2} \cdot \pF^\pm_{j+k}, \ s = 1, \cdots, 5,
\end{equation}
where $\pF^\pm$ is given by \er{F-pm-LF} below. Applying the scalar WENO scheme \er{2.10-96} to each of the characteristic field, we have for $s$-th field of positive flux,
\begin{equation*}
 \hat F^+_{j+1/2, s} = \sum_{k=0}^{3} \om_{k, s} q_k^+(F^+_{j+k-2, s}, \cdots, F^+_{j+k, s}).
\end{equation*}
Details about this reconstruction can be found in \S~\ref{sec.2.2}. Finally, the numerical fluxes obtained in each characteristic field can then be projected back to the physical space by
\begin{equation}\lb{fs-1d-2}
 \hat \pF_{j+1/2} = \sum_{s=1}^{5} \hat F_{j+1/2, s} \pr_{s, j+1/2}.
\end{equation}

\subsection{Positivity preserving scheme}\label{sec.positive}

One important issue of high-order conservative schemes is that nonphysical negative density or pressure (failure of positivity) can lead to an ill-posed system, which may cause blow-up of the numerical scheme. In particular, for high-order schemes positivity failure can occur due to interpolation errors at or near very strong discontinuities even though the flow physically is far away from vacuum. Thus, schemes which can preserve the positivity of density and pressure are very desiring. Along this direction, various bound-preserving limiters have been proposed, which have been reviewed by \citet{Shu2016}.

Recently, \citet{Hu2013} proposed an \textit{a posteriori} approach within the content of FDM, which first detects the critical numerical fluxes that may lead to negative pressure or density and then imposes a positivity-preserving flux limiter by combining the high-order numerical flux with the 1st-order LF flux to satisfy a sufficient condition for preserving positivity to correct the fluxes. This approach is simple and has a strong mathematical foundation. However, it can not be directly implemented in our scheme yet since the pressure or density is set to a prescribed small value when negative pressure or density occurs, which is reasonable only for nearly vacuum flow but will lead to the wrong solution in shocked flow where negative pressure or density occurs due to nonphysical oscillations. Recall that all shock capturing schemes at the location of shock waves are no more than 1st-order, we may set the flux limiter to be zero instead of to set the pressure or density to be a prescribed small value when negative pressure or density occurs. Combined with special initial conditions and the improvement to be discussed later, this technique can indeed make our scheme suitable for the simulation of CIT even with $M_t = 3.0$. However, our numerical results (not shown here) indicate that the direct interchange between the 7th-order WENO scheme and the 1st-order monotone scheme will lead to excessive numerical viscosity. Thus, we prefer to only utilize the idea of \citet{Hu2013} as the RFD criterion of the ROR-WENO method.

Following \citet{Hu2013}, the general explicit $k$th-order conservative scheme with Euler-forward time integration can be written as
\begin{equation}\label{HAS-7-3d}
  \begin{split}
  \pU_{i, j, k}^{n+1} = & - \fr{\Delta t}{\Delta x} (\hat \pF_{i+\fr{1}{2}, j, k} - \hat \pF_{i-\fr{1}{2}, j, k} ) \\
  & - \fr{\Delta t}{\Delta y}  (\hat \pG_{i, j+\fr{1}{2}, k} - \hat \pG_{i, j-\fr{1}{2}, k} ) \\
  & - \fr{\Delta t}{\Delta z}  (\hat \pH_{i, j, k+\fr{1}{2}} - \hat \pH_{i, j, k-\fr{1}{2}} ).
  \end{split}
\end{equation}
The positivity-preserving property for the scheme \er{HAS-7-3d} refers to the property that the density and pressure are positive for $\pU_{i, j, k}^{n+1}$ when $\pU_{i, j, k}^n$ has positive density and pressure. For convenience, we rewrite \er{HAS-7-3d} as
\begin{equation}\label{HAS-7-3d-1}
  \begin{split}
  \pU_{i, j, k}^{n+1} = & \ \ \ \ \fr{1}{6} \left( \pU_{i, j, k}^n - 6 \fr{\Delta t}{\Delta x} \hat \pF_{i+\fr{1}{2}, j, k} \right) + \fr{1}{6} \left( \pU_{i, j, k}^n+ 6 \fr{\Delta t}{\Delta x} \hat \pF_{i-\fr{1}{2}, j, k} \right) \\
  & + \fr{1}{6} \left( \pU_{i, j, k}^n  - 6 \fr{\Delta t}{\Delta y} \hat \pG_{i, j+\fr{1}{2}, k} \right) + \fr{1}{6} \left( \pU_{i, j, k}^n + 6 \fr{\Delta t}{\Delta y} \hat \pG_{i, j-\fr{1}{2}, k} \right) \\
  & + \fr{1}{6} \left( \pU_{i, j, k}^n  - 6 \fr{ \Delta t}{\Delta z} \hat \pH_{i, j, k+\fr{1}{2}} \right) + \fr{1}{6} \left( \pU_{i, j, k}^n + 6 \fr{ \Delta t}{\Delta z} \hat \pH_{i, j, k-\fr{1}{2}} \right).
  \end{split}
\end{equation}
Thus, a \textit{sufficient} condition for preserving positivity is that all terms within parentheses have positive density and pressure.
For instance, if $\pU_{i, j, k}^{\pm} = \pU_{i, j, k}^n \mp (6 \Delta t/\Delta x) \hat \pF_{i\pm\fr{1}{2}, j, k}$, then the positivity-preserving property means that $\rho(\pU_{i, j, k}^\pm) > 0$ and $p(\pU_{i, j, k}^\pm) >0$.

To illustrate how the idea of positivity preserving works as the RFD criterion of the ROR-WENO method, we may take the $x$-component as an example. First, define
\begin{equation}
  \left\{ \begin{split}
  & \pU_{i, j, k}^{+} = \pU_{i, j, k}^n - 6 \fr{\Delta t}{\Delta x} \pF^{\rm WENO}_{i+\fr{1}{2}, j, k}, \\
  & \pU_{i+1, j, k}^{-} = \pU_{i+1, j, k}^n + 6 \fr{\Delta t}{\Delta x} \pF^{\rm WENO}_{i+\fr{1}{2}, j, k},
  \end{split} \right.
\end{equation}
where the numerical flux $\hat \pF_{i+\fr{1}{2}, j, k}$ has been replaced by that of a high-order WENO scheme $\pF^{\rm WENO}_{i+\fr{1}{2}, j, k}$. Next, calculate the corresponding density and pressure of $\pU_{i, j, k}^{+}$ and $\pU_{i+1, j, k}^{-}$. If either of them is negative or smaller than a prescribed value, then we reduce the order of the WENO scheme and recompute the WENO flux. This can be recursively performed until the 1st-order global LF flux
\begin{equation}\label{HAS-12}
  \hat \pF_{i + \fr{1}{2}, j, k}^{\rm LF} = \fr{1}{2} [ \pF_{i, j, k} + \pF_{i+1, j, k} + \hat \lambda_s (\pU_{i, j, k}^n - \pU_{i+1, j, k}^n)],
\end{equation}
which has the positivity-preserving property under a proper CFL number, where $\hat \lambda_s$ is given by \er{F-pm-c-3} below. Another good property of this procedure is that it can be performed on an arbitrary single point. Similar procedures can be applied to the flux $\pG$ and $\pH$, which are omitted here.

\subsection{Upwind-capturing methods}\label{sec.upwind}

When approximating a hyperbolic system of conservation laws with the so-called upwind differences, we must, in the first place, determine in which direction each of a variety of signals moves through the computational grid. For this purpose, two physical models of the interaction between computational cells can be used. The first one is the Riemann or flux difference splitting (FDS) approach, where neighboring cells interact through discrete, finite-amplitude waves, which are found by solving Riemann's initial-value problem for the discontinuity at the cell interface. Examples are the methods of \citet{Roe1981} and of \citet{Osher1981}. The other one is the Boltzmann or FVS approach, where the interaction of neighboring cells is accomplished through mixing of pseudo-particles that move in and out of each cell according to a given velocity distribution. Examples are the methods of \citet{Steger1981} and of \citet{Leer1982}. Since this approach is relatively simple, it is adopted in our new scheme as well as that of \citet{Wang2010}.

In our numerical tests, we found that all schemes even the 1st-order upwind scheme based on local LF flux or SW flux fails when $M_t > 1$. This abnormal phenomenon makes us to take a deep look at the FVS approach, which can be understood more clearly by tracing back to the ``beam scheme'' of \citet{Sanders1974}, where the velocity distribution function was represented by three delta-function with four unknowns, of which only one is a free parameter when the mass, momentum, and kinetic energy are preserved. When the free parameter is properly chosen, the SW flux is recovered. Realizing this close relation between the ``beam scheme'' and SW flux, we think the failure of the latter at $M_t>1$ cases is due to the following conjecture: The local instantaneous velocity distribution function is very close to that of the global averaged one when there is no very strong shock waves in the flow, say, $M_t < 1$; however, when $M_t > 1$ the shocklets can merge to very strong shock waves, leading the local instantaneous velocity distribution function very different to that of the global averaged one. To distinguish the forward- and backward-moving particles, in our viewpoint, the averaged velocity distribution function rather than the local instantaneous one should be used. Thus, we propose to use the global LF flux as the building blocks of the WENO scheme. Take the flux $\pF$ along the $x$ direction as an example again, there is
\begin{equation}\label{F-pm-LF}
  \pF = \pF^+ + \pF^-, \quad
  \pF^\pm = \fr{1}{2} (\pF \pm \hat \lambda_s \pU),
\end{equation}
where
\begin{equation}\label{F-pm-c-3}
  \hat \lambda_s = \max_{1\le j \le N} |\lambda_{s; j}|, \ s = 1, \cdots, 5,
\end{equation}
with $\lambda_{s}$ given by \er{lambda-A} and $N$ representing the number of points involved in the WENO flux. With this flux as building blocks, the 1st-order scheme is indeed very robust.

\subsection{Cooling function}\label{sec.cooling}

The cooling function of \citet{Wang2010} is given by \er{38}, which is accomplished in program by changing the local internal energy per unit volume directly but keeping the local density and the average internal energy unchanged after cooling. Denote $E_0$ and $E_1$ as the local internal energy before and after cooling, respectively, there is
\begin{equation}\label{38-1}
  E_1 - E_0 = a E_0^b.
\end{equation}
Note that we have substituted $T$ in \er{38} by $E_0$ since $E_0$ is proportional to $T_0$ and $\rho_0$, of which the latter can be absorbed into the factor $a$.

Denote $\overline{E_0}$ and $\overline{E_1}$ as the averaged internal energy before and after cooling, respectively. Since the averaged internal energy must be kept as constant, there is $\overline{E_1} = \rm const.$, which is the averaged internal energy of the system in stationary state. Let $b$ be the adjustable parameter, then \er{38-1} can be reduced to
\begin{equation}\label{38-2}
  E_1 = E_0 + \fr{\overline{E_1} - \overline{E_0}}{\overline{E_0^b}} E_0^b.
\end{equation}
Now it can be seen clearly when negative temperature will happen: If and only if $\overline{E_0} > \overline{E_1}$, i.e., heat is removed from the system, then negative temperature may appear. In contrast, however, if $\overline{E_0} < \overline{E_1}$, i.e., heat is added into the system, then no negative temperature will present when $E_0 > 0$.

It is hard to obtain all possible values of $b$ such that $E_1$ is positivity preserving, that is, $E_1 > 0$ as long as $E_0>0$. This difficulty comes from the fact that $E_0$ is a random number such that its operators of expectation and $b$th-order power can not be interchanged arbitrarily. Nevertheless, there are indeed some cases that they can be interchanged, say, $b = 0$ and $b=1$. The former is the uniform cooling as adopted by \citet{Wang2010}, which indicates a translation of the reference internal energy. Since $E_0$ may have very large fluctuations at very large $M_t$, this translation may lead some value of $E_0$ falling below zero, resulting simulation blow-up. For the latter, \er{38-2} reduces to
\begin{equation}\label{38-3}
  E_1 = \fr{\overline{E_1}}{\overline{E_0}} E_0,
\end{equation}
which means that the amount of heat removed or added is not a constant but a value proportional to the local internal energy and the inverse of averaged internal energy before cooling. Obviously, this operation is positivity preserving and keeps the averaged internal energy constant. Thus, we still adopt the cooling function \er{38}, but propose the value of the parameter $b$ to be 1 rather than 0, 2, or 4 as \citet{Wang2010} did. Note that this difference also has a negligible effect on the statistical properties of the flow field at stationary state for $M_t < 1$, which has been confirmed by our numerical results (see \S~\ref{sec.4.2} below).

\section{Numerical results}\label{sec.4}

In this section, we present several numerical results to demonstrate the capabilities of our hybrid method, especially its improvements and extension of Wang's scheme. All problems are numerically solved by the 2nd-order TVD Runge-Kutta time advanced method \citep{Shu1988}.

\subsection{One-dimensional problems}\label{sec.4.1}

\begin{figure}
  \centering
  \includegraphics[width=0.49\textwidth]{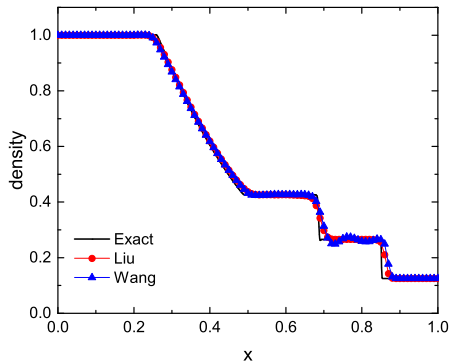}\\
  \caption{Numerical results of the Sod problem at $t=0.2$. Exact: Liu's scheme with grid number $N=1001$; Liu: Liu's scheme with $N=101$; Wang: Wang's scheme with $N=101$.} \label{fig.Sod}
\end{figure}

\begin{figure}
  \centering
  \includegraphics[width=0.49\textwidth]{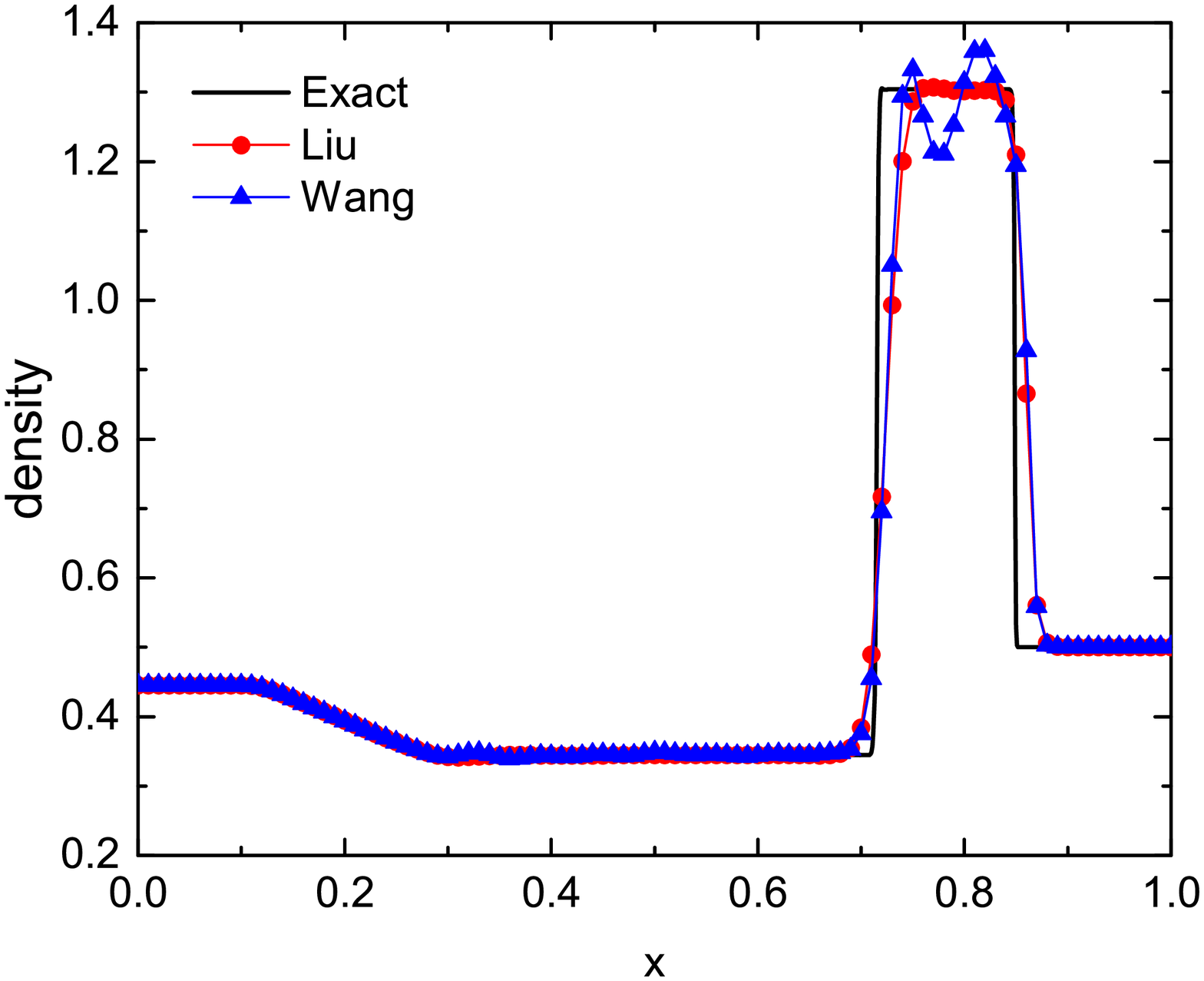}\\
  \caption{Numerical results of the Lax problem at $t=0.14$. Exact: Liu's scheme with grid number $N=1001$; Liu: Liu's scheme with $N=101$; Wang: Wang's scheme with $N=101$.} \label{fig.Lax}
\end{figure}

\begin{figure}
  \centering
  \includegraphics[width=0.49\textwidth]{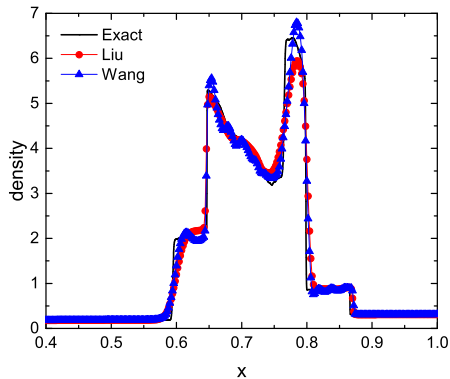}\\
  \caption{Numerical results of the Woodward-Colella problem at $t=0.038$. Exact: Liu's scheme with grid number $N=4001$; Liu: Liu's scheme with $N=501$; Wang: Wang's scheme with $N=501$.}\label{fig.IBWP}
\end{figure}

In this subsection we will consider three classical problems, namely, the Sod problem, the Lax problem, and the Woodward-Colella problem. In particular, the initial condition for the Sod problem is \citep{Sod1978}
\begin{equation}\label{eq.Sod}
  (\rho, u, p) = \left\{
  \begin{split}
  &(1.000, 0, 1.0), & \quad & x < 0.5, \\
  &(0.125, 0, 0.1), & \quad & x \ge 0.5, \\
  \end{split}
  \right.
\end{equation}
and the final simulation time is $t=0.2$; the initial condition for the Lax problem is \citep{Lax1954}
\begin{equation}\label{eq.Lax}
  (\rho, u, p) = \left\{
  \begin{split}
  &(0.445, 0.689, 3.528), & \quad & x < 0.5, \\
  &(0.500, 0.000, 0.571), & \quad & x \ge 0.5, \\
  \end{split}
  \right.
\end{equation}
and the final simulation time is $t=0.14$; and the initial condition for the Woodward-Colella problem is \citep{Woodward1984}
\begin{equation}\label{IBWP}
  (\rho, u, p) = \left\{
  \begin{split}
  &(1, 0, 1000), & \quad 0.0 \le x < 0.1, \\
  &(1, 0, 0.01), & \quad 0.1 \le x < 0.9, \\
  &(1, 0, 100), & \quad 0.9 \le x \le 1.0, \\
  \end{split}
  \right.
\end{equation}
and the boundary conditions are reflected boundaries at $x=0$ and $1$. These problems are numerically solved by both the present scheme (hereafter we call it Liu's scheme for short) and Wang's scheme \citep{Wang2010}. The results are shown in Figs.~\ref{fig.Sod}, \ref{fig.Lax}  and \ref{fig.IBWP}, where the exact solutions are obtained by Liu's scheme with a much larger grid number. Note that in these simulations only the shock-capturing sub-scheme is activated.

As seen from these figures, the numerical solution of the Sod problem from Wang's scheme has already shown some spurious oscillations even though the strengths of discontinuities are still very weak (see Fig.~\ref{fig.Sod}), which is more significant when the strengths become strong (see Figs.~\ref{fig.Lax} and \ref{fig.IBWP}). This is a typical characteristic of high-order shock-capturing schemes, of which the oscillations can be depressed if low-order schemes are used (not shown here). In contrast, the results obtained from Liu's scheme show almost none oscillations in all these three problems, indicating that characteristic-wise reconstruction alone can indeed depress spurious oscillations when the discontinuities are not too strong (Figs.~\ref{fig.Sod} and \ref{fig.Lax}). As a result, we can simply turn off the ROR procedures. However, this is not the case for the Woodward-Colella problem, where the strengths of discontinuities are too strong such that characteristic-wise reconstruction alone still leads to blow-up of the simulation with the 7th-order WENO. This can be overcome either by lower the order of the scheme or turn on the ROR procedures as given in \S~\ref{sec.positive}. Thus, Wang's scheme has indeed been improved by Liu's scheme.

\subsection{Forced compressible turbulence}\label{sec.4.2}

In this subsection we shall consider a few simulations for forced compressible turbulence at different resolutions and schemes, as well as different turbulent Mach numbers.

\subsubsection{Comparison with Wang's scheme}

Now we shall consider a few simulations for forced compressible turbulence at $128^3$ resolutions for both Wang's and Liu's scheme, where the velocity field is initialized using a random field with a prescribed energy spectrum and the normalized temperature and density are simply initialized to one at all spatial points. The velocity field is forced by fixing the total kinetic energy per unit mass in the first two wavenumber shells to $E(1) = 1.242477$ and $E(2) = 0.391356$ and the forcing field is made incompressible. The hyperviscosity coefficient is $\nu_n = 0.05$ and the dimensionless time step is $\Delta t = 0.001$. The total computation time is $t=20$ and the typical eddy turnover time is $T_e \approx 1.2$. The statistical quantities are averaged over the time interval $8 \lesssim t/T_e \lesssim 16$, where the flow has reached its statistically stationary state. More computation details can be found in \citet{Wang2010}.

By setting the reference Reynolds number $Re=200$ and Mach number $M=0.35$, which come from the nondimensionalization of the governing equation \er{NS-3d}, we obtain an average turbulent Mach number $M_t \approx 0.8$ and Taylor microscale Reynolds number $R_\lambda \approx 120$ for the first case as shown in the second block of Table~\ref{tab1}. While for the second case as shown in the third block of Table~\ref{tab1}, we obtain $M_t \approx 1.0$ and $R_\lambda \approx 120$ by setting $M=0.45$ and $Re=200$. Other statistic parameters are also summarized in Table~\ref{tab1}. Evidently, these statistic quantities are all consistent with each other perfectly, indicating the correctness of our scheme.

\begin{table}
  \centering
  \caption{Flow statistics of stationary compressible isotropic turbulence obtained by different schemes}
  \begin{tabular}{|cccccccccc|}  \hline
  Scheme & $M_t$ & $R_\lambda$& $\varepsilon$ & $\eta$ & $L_f$ & $T_e$ & $\theta_{\rm rms}$ & $\omega_{\rm rms}$ & $S_3$ \\ \hline
  Wang & 0.79 & 118 & 0.59 & 0.022 & 1.52 & 1.18 & 2.40 & 10.4 & $-0.55$ \\
  Liu  & 0.80 & 115 & 0.65 & 0.021 & 1.50 & 1.14 & 2.34 & 11.0 & $-0.50$ \\  \hline
  Wang & 1.02 & 116 & 0.58 & 0.022 & 1.52 & 1.18 & 3.54 & 9.92 & $-0.85$ \\
  Liu  & 1.01 & 118 & 0.55 & 0.022 & 1.53 & 1.20 & 3.19 & 9.78 & $-0.74$ \\  \hline
  \end{tabular}
  \label{tab1}
\end{table}

\begin{figure}
  \centering
  \subfigure[$M_t \approx 0.8$]{\includegraphics[width=0.49\textwidth]{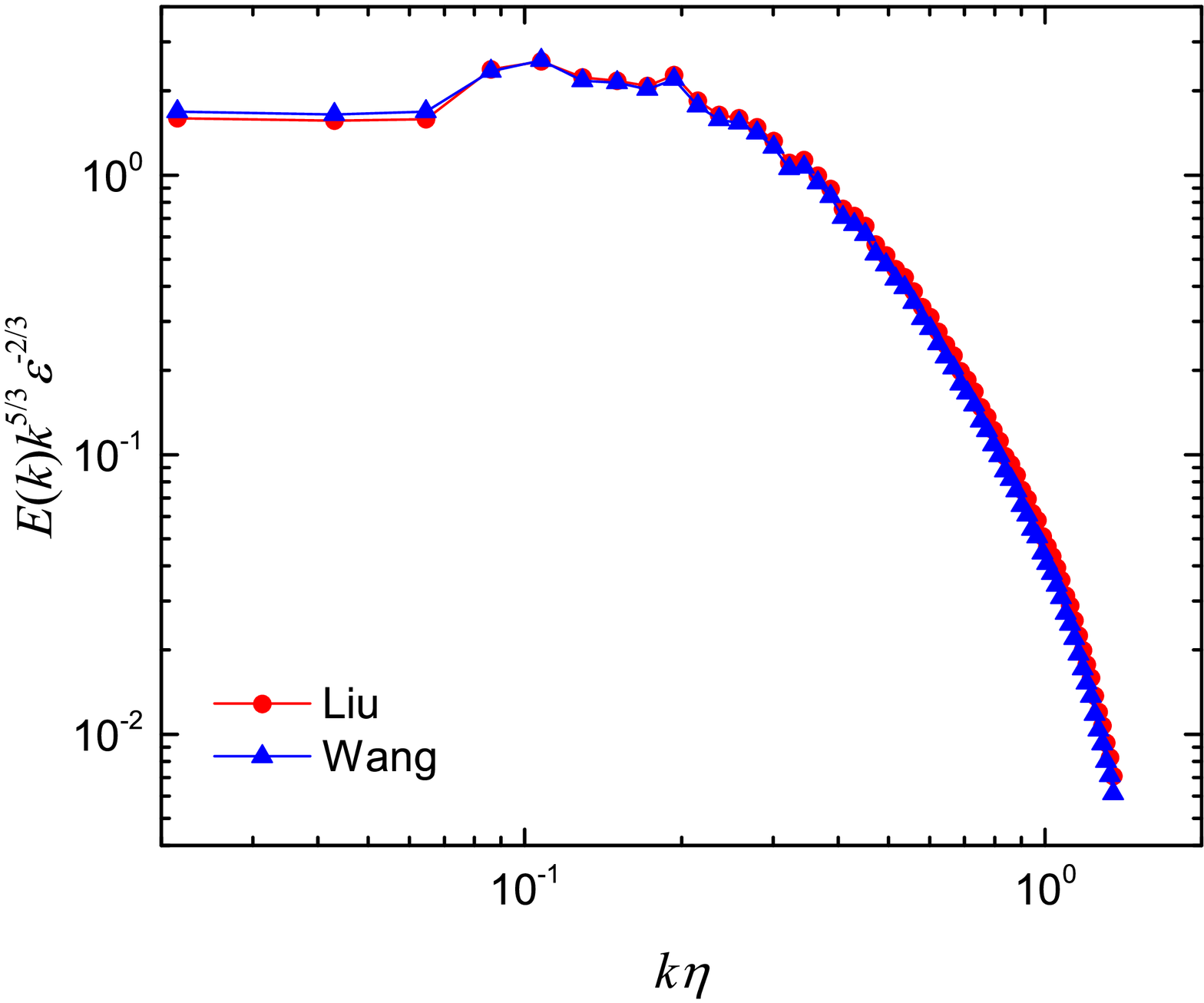}}
  \subfigure[$M_t \approx 1.0$]{\includegraphics[width=0.49\textwidth]{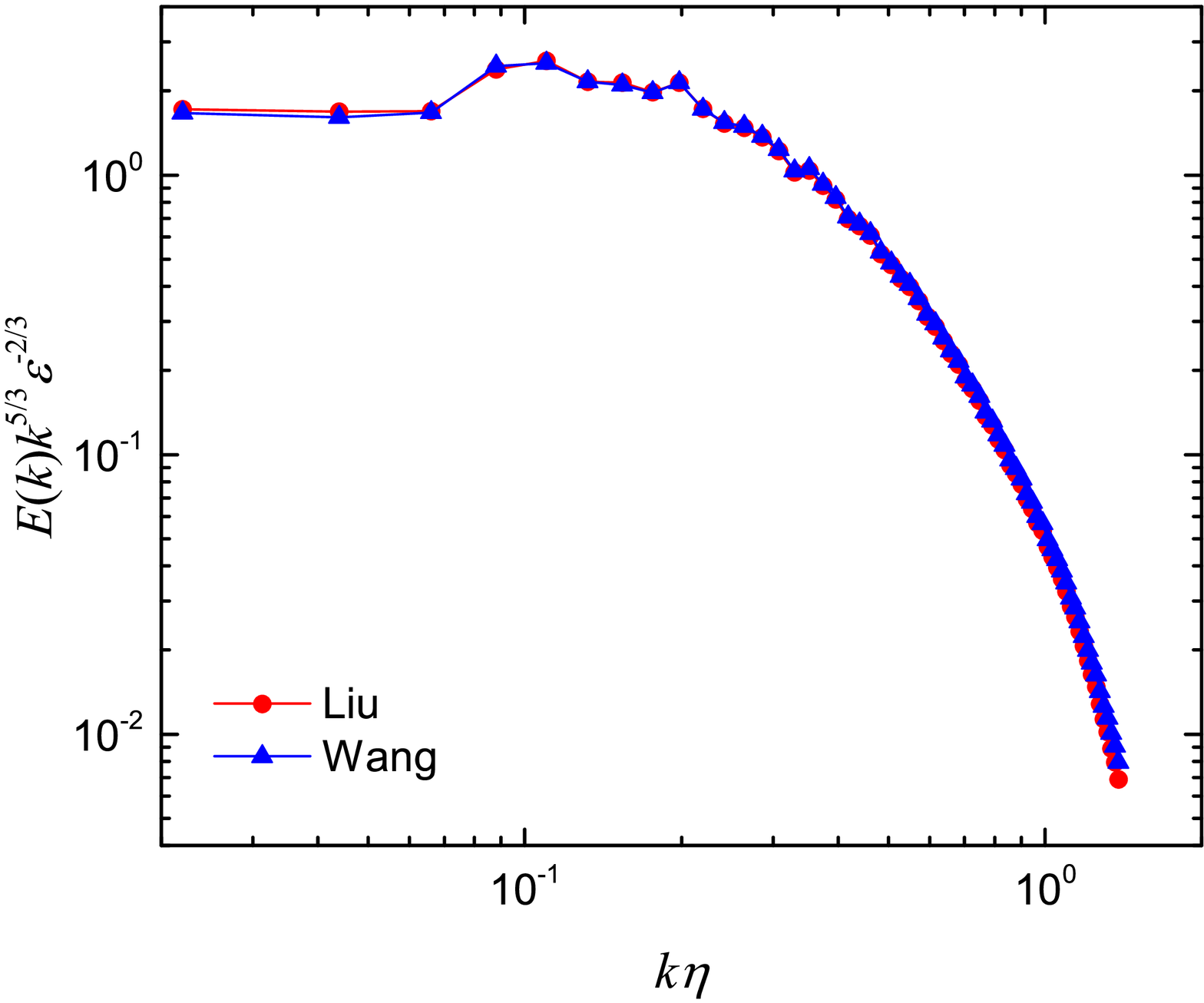}}
  \caption{Comparison of averaged compensated kinetic energy spectra per unit volume of stationary compressible isotropic turbulence with $R_\lambda \approx 120$, simulated with $128^3$ grid resolution}
  \label{fig.EM07}
\end{figure}

\begin{figure}
  \centering
  \subfigure[$M_t \approx 0.8$]{\includegraphics[width=0.49\textwidth]{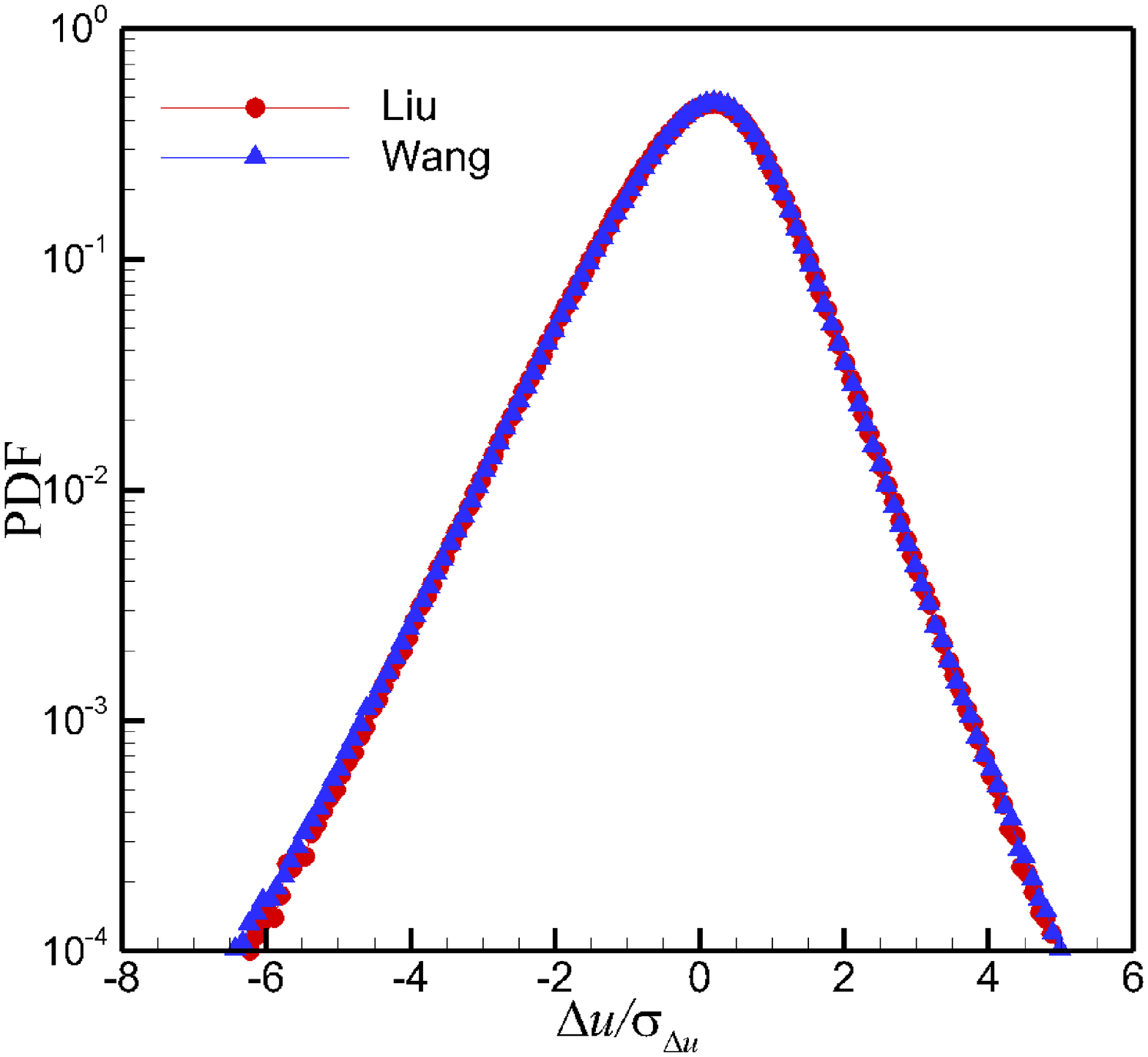}}
  \subfigure[$M_t \approx 1.0$]{\includegraphics[width=0.49\textwidth]{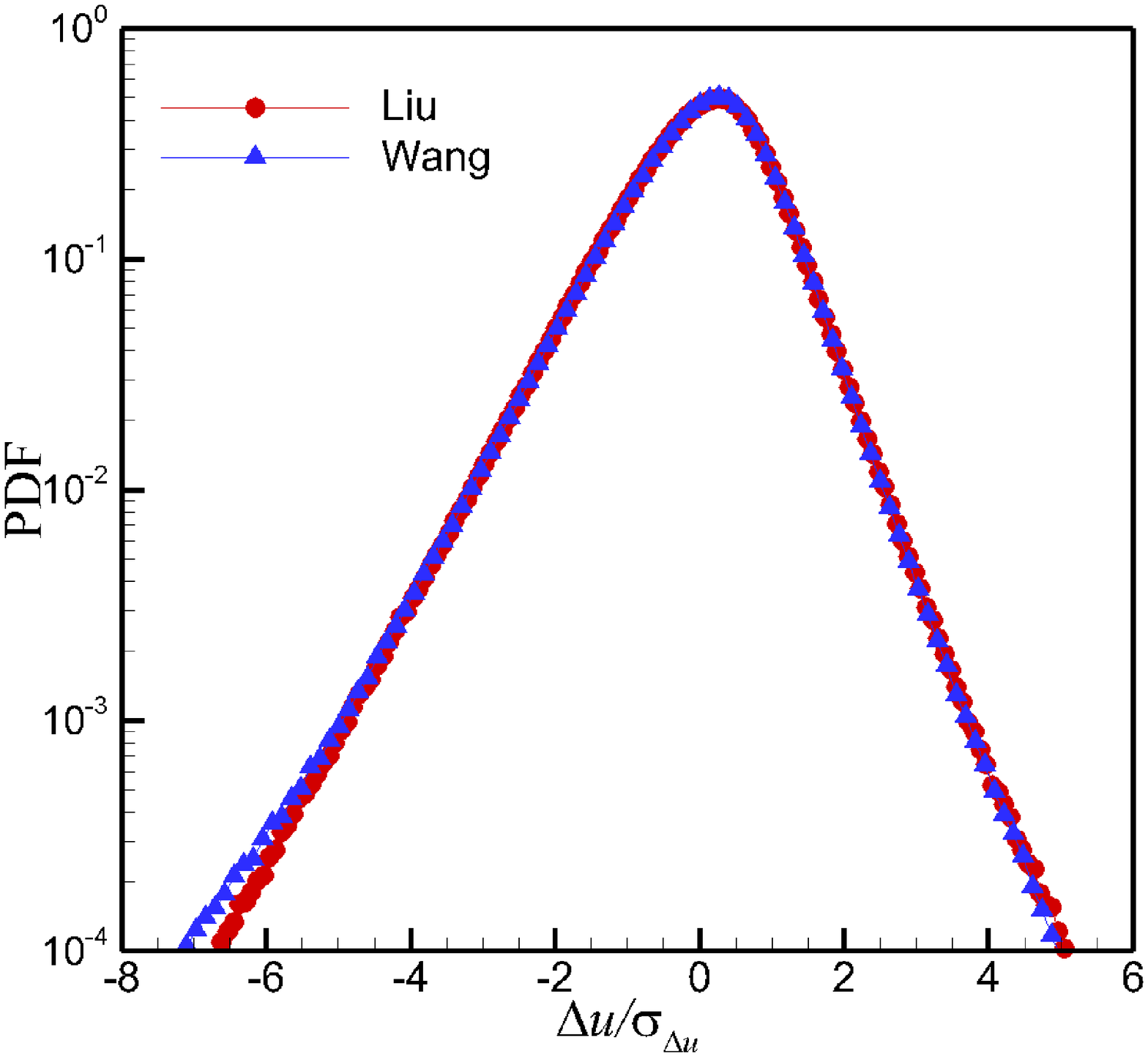}}
  \caption{Comparison of averaged PDFs of longitudinal velocity increment at a separation equal to $\Delta x$ of stationary compressible isotropic turbulence with $R_\lambda \approx 120$, simulated with $128^3$ grid resolution, where $\sigma_{\Delta u}$ denotes the standard deviation of $\Delta u$}
  \label{fig.velM07}
\end{figure}

\begin{figure}
  \centering
  \subfigure[$M_t \approx 0.8$]{\includegraphics[width=0.49\textwidth]{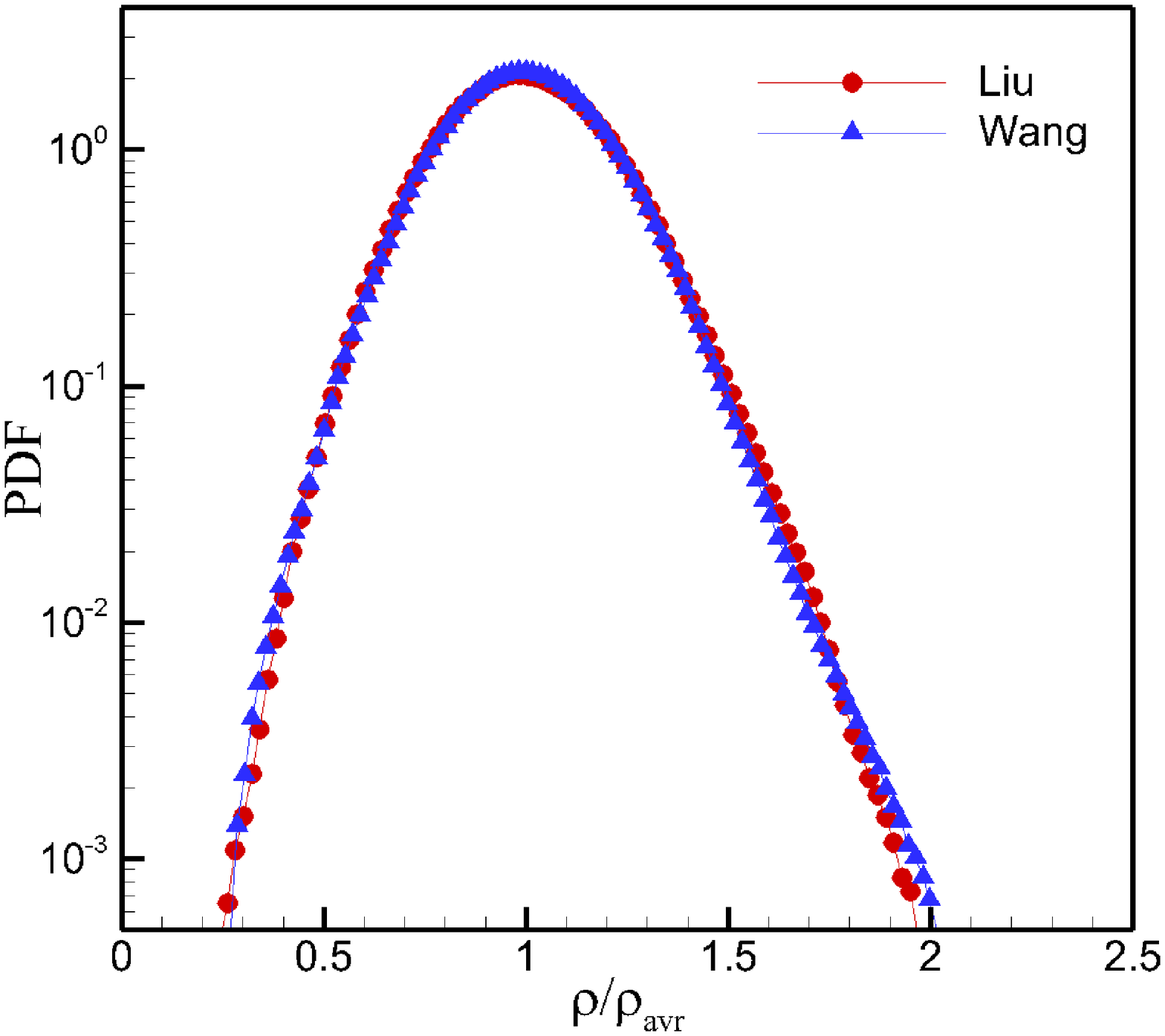}}
  \subfigure[$M_t \approx 1.0$]{\includegraphics[width=0.49\textwidth]{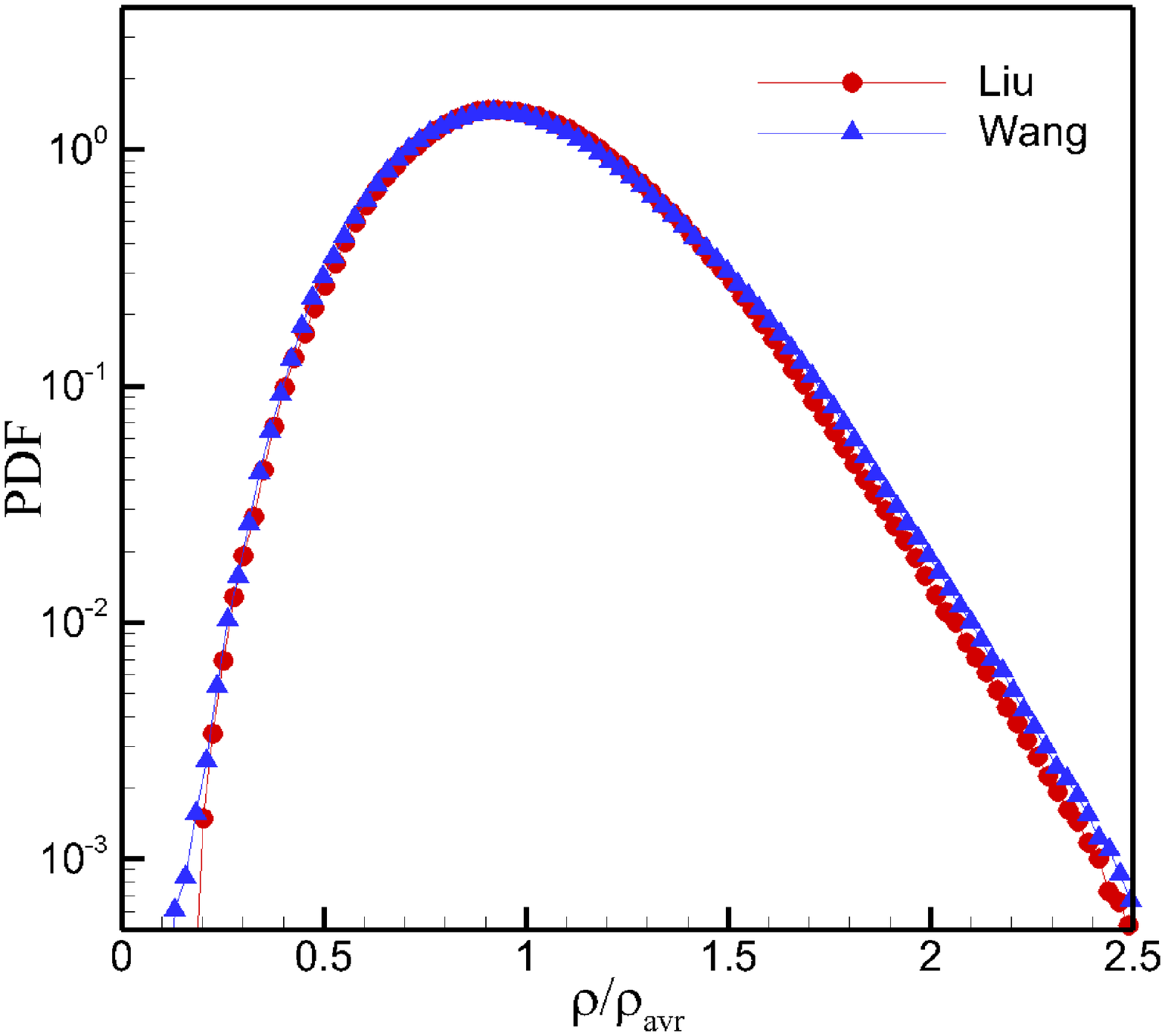}}
  \caption{Comparison of averaged PDFs of local density of stationary compressible isotropic turbulence with $R_\lambda \approx 120$, simulated with $128^3$ grid resolution, where $\rho_{\rm avr}$ denotes the averaged density}
  \label{fig.rhoM07}
\end{figure}

Fig.~\ref{fig.EM07} compares the averaged compensated kinetic energy spectra per unit volume obtained by Wang's and Liu's schemes, where $\varepsilon$ is the viscous dissipation rate, $\eta$ is the Kolmogorov length scale, and $E(k)$ is the kinetic energy spectra per unit volume. The results obtained by these two schemes are also consistent with each other. Similar conclusions can further be obtained for other quantities, for example, the probability density functions (PDFs) of longitudinal velocity increment and local density as shown in Figs.~\ref{fig.velM07} and \ref{fig.rhoM07}, respectively, where, in order to display the PDF tails more clearly, the logarithmic coordinate has been used. Note that although the energy spectrum and the PDF of velocity increment have a negligible difference for these two flows, the PDF of local density has a strong dependence on the turbulent Mach number. Nevertheless, the consistence of the results obtained by these two schemes indicates that the good properties of Wang's scheme has indeed been inherited by Liu's scheme.

\subsubsection{Dependence on the grid resolution}

Three grid resolutions ($128^3$, $256^3$ and $512^3$) are considered in order to assess any dependence of small-scale flow statistics on the grid resolution. After reaching the statistically stationary state, the time period of $10 \le  t \le 20$ is used to obtain statistically averages of interested quantities. By setting $M=0.45, 0.90$ and $Re=200$, we obtain $M_t \approx 1.02, 2.06$ and $R_\lambda \approx 106$ for sufficient resolution, respectively. Note that the initial conditions for $M = 0.45$ with grid resolutions $128^3$ and $256^3$ are the same with random velocity field and uniform thermodynamic field, while the others are the statistically stationary forced compressible turbulence with a relative smaller $M$.

Other flow statistics of the simulations are compiled in Table~\ref{tab2} for three grid resolutions. The resolution parameters $k_{\rm max} \eta$ are, respectively, 0.92, 1.87, and 3.75 for $M = 0.45$, where Kolmogorov length scales $\eta$ are all around $0.022$ and the largest wavenumbers $k_{\rm max} = N/3$ are, respectively, 42, 85, and 170. Similar results can be obtained for the case $M = 0.90$. The statistics shown in Table~\ref{tab2} imply that the small-scale flow is already well resolved in the $256^3$ simulation since $1.87 > \pi/2 = 1.57$, needless to say the $512^3$ simulation. In addition, if we set $Re=100$ then all these three resolutions can well resolve the small-scale flow (results not shown here).

\begin{table}
  \centering
  \caption{Flow statistics of stationary compressible isotropic turbulence obtained by different resolutions}
  \begin{tabular}{|crcccccccc|} \hline
  Resolution & $M_t$ & $R_\lambda$& $\varepsilon$ & $\eta$ & $L_f$ & $T_e$ & $\theta_{\rm rms}$ & $\omega_{\rm rms}$ & $S_3$ \\ \hline
  $128^3$ & 1.01 & 118 & 0.55 & 0.022 & 1.53 & 1.20 & 3.19 & 9.78 & $-0.74$ \\
  $256^3$ & 1.01 & 111 & 0.59 & 0.022 & 1.53 & 1.19 & 3.76 & 9.92 & $-1.26$ \\
  $512^3$ & 1.02 & 106 & 0.63 & 0.022 & 1.52 & 1.19 & 4.36 & 9.95 & $-2.56$ \\ \hline
  $128^3$ & 1.92 & 143 & 0.32 & 0.030 & 1.63 & 1.36 & 3.38 & 6.50 & $-1.57$ \\
  $256^3$ & 2.00 & 118 & 0.41 & 0.026 & 1.61 & 1.34 & 4.90 & 6.74 & $-3.03$ \\
  $512^3$ & 2.06 & 106 & 0.53 & 0.025 & 1.56 & 1.29 & 5.97 & 7.20 & $-4.74$ \\ \hline
  \end{tabular}
  \label{tab2}
\end{table}

\begin{figure}
  \centering
  \subfigure[$M_t \approx 1.0$]{\includegraphics[width=0.49\textwidth]{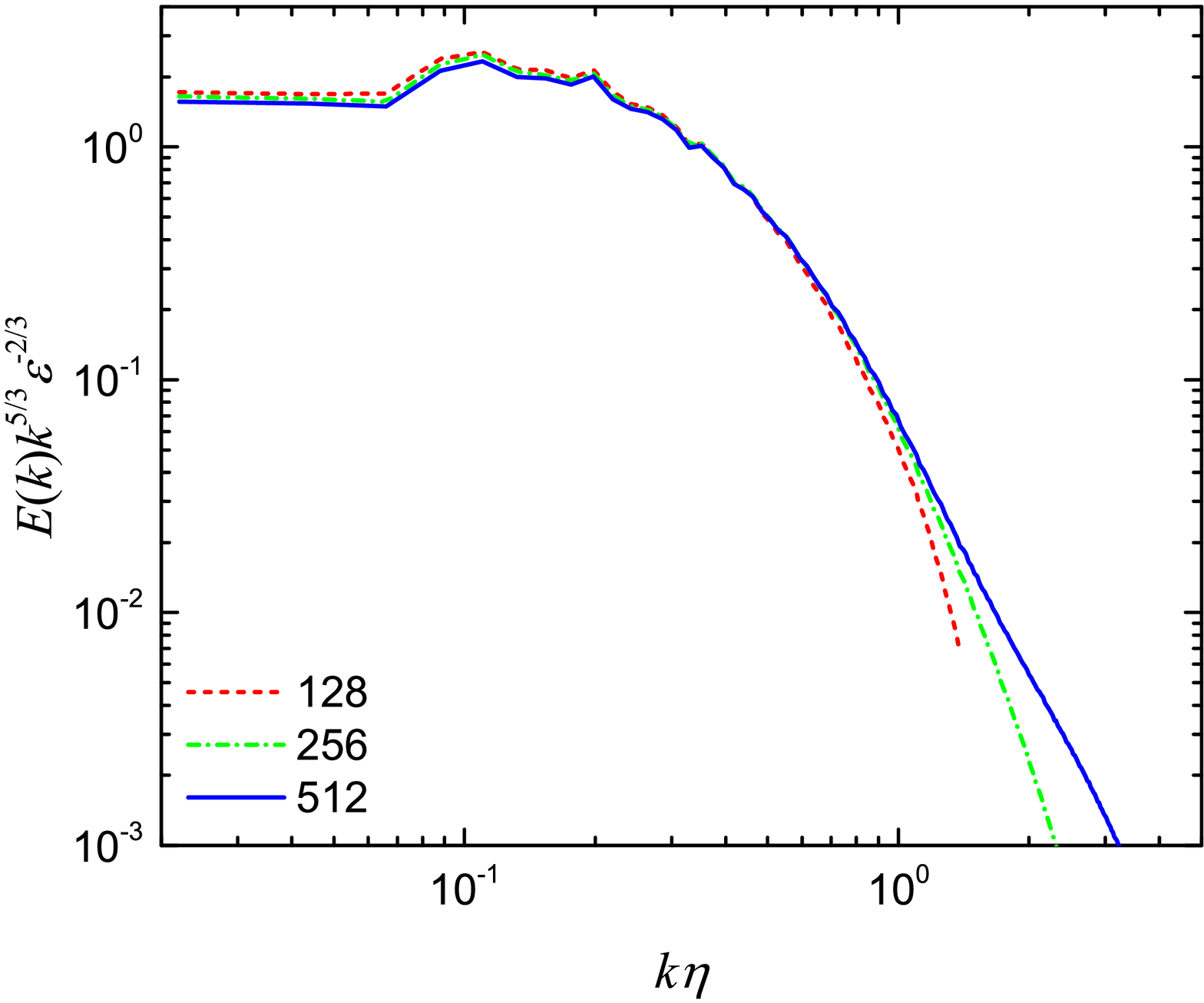}}
  \subfigure[$M_t \approx 2.0$]{\includegraphics[width=0.49\textwidth]{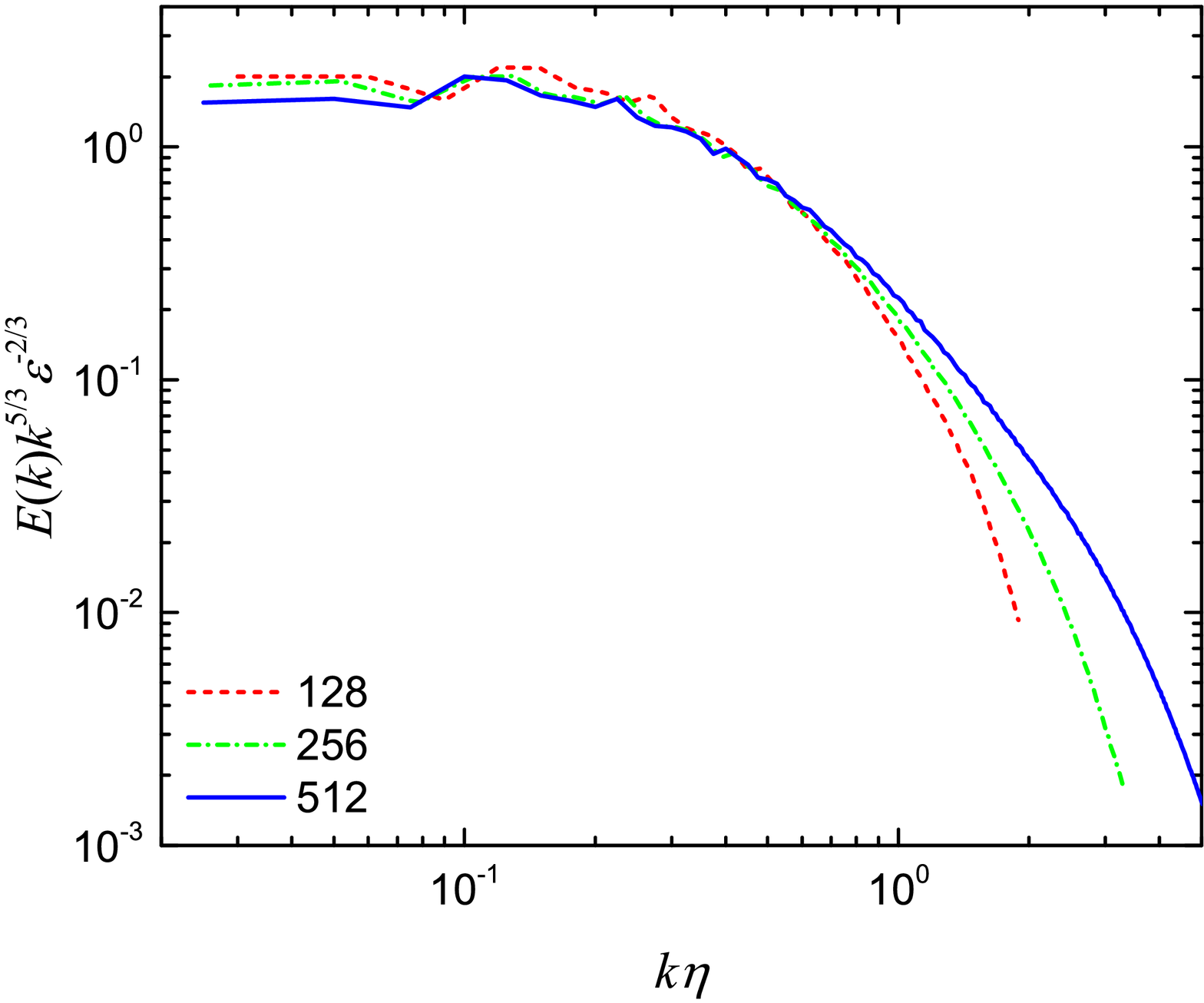}}
  \caption{Grid convergence of averaged compensated kinetic energy spectra per unit volume of stationary compressible isotropic turbulence with $R_\lambda \approx 110$}
  \label{fig.M090R200}
\end{figure}

In Fig.~\ref{fig.M090R200} we plot the averaged compensated kinetic energy spectra per unit volume averaged over the time interval $8 \lesssim t/T_e \lesssim 16$ for different grid resolutions and Mach numbers. All spectra tend to converge to that of $512^3$ resolution. In particular, the energy spectra from $256^3$ and $512^3$ resolutions overlap in almost all resolved scale ranges, implying the convergence of energy spectra under this grid refinement. In addition, a short inertial range is identified, with a Kolmogrov constant of about $E(1) \varepsilon^{-2/3} \approx 1.6$. Evidently, this inertial range will extend when Reynolds number increases since $\eta$ decreases with $Re$.

\begin{figure}
  \centering
  \subfigure[$M_t \approx 1.0$]{\includegraphics[width=0.49\textwidth]{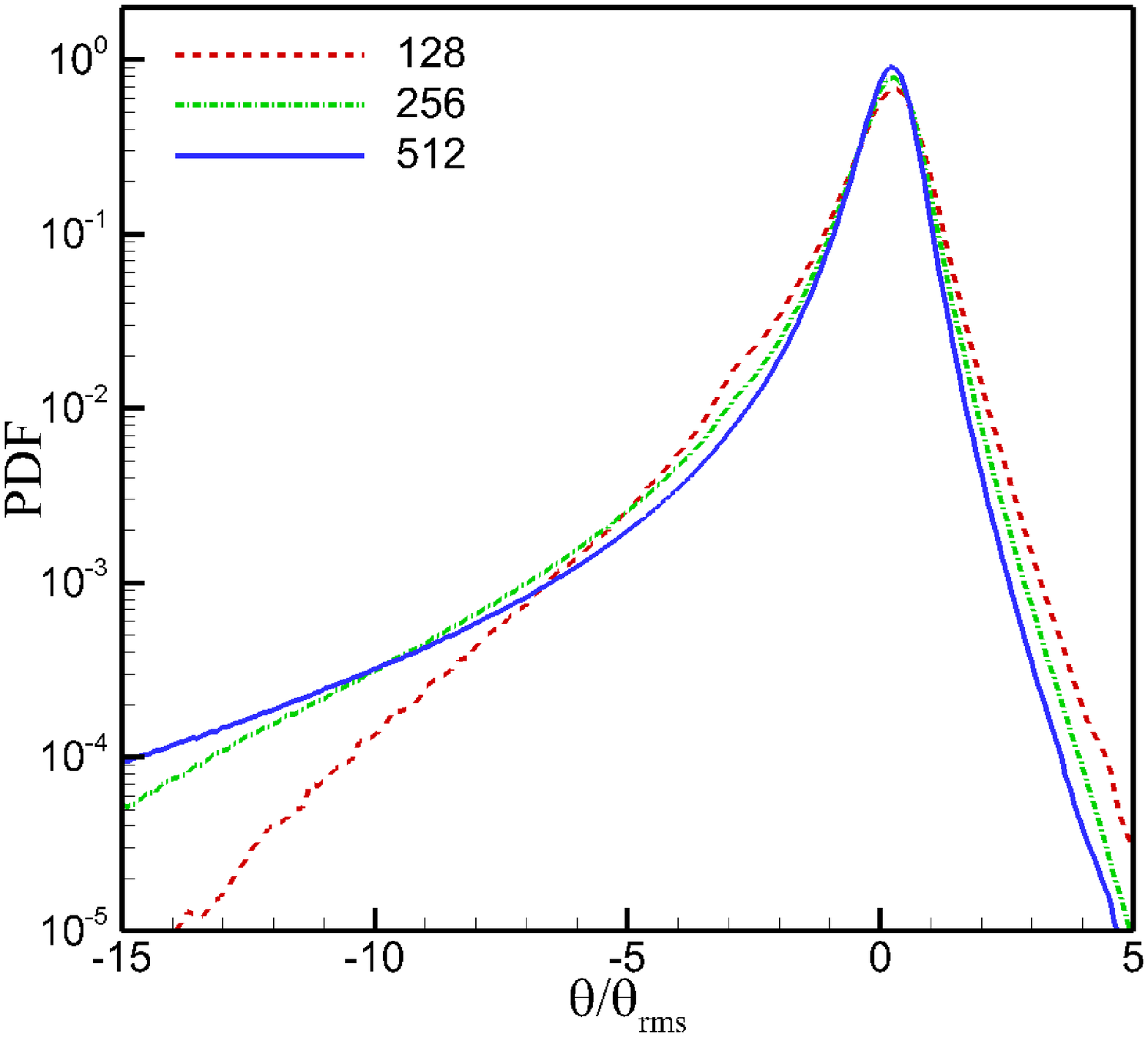}}
  \subfigure[$M_t \approx 2.0$]{\includegraphics[width=0.49\textwidth]{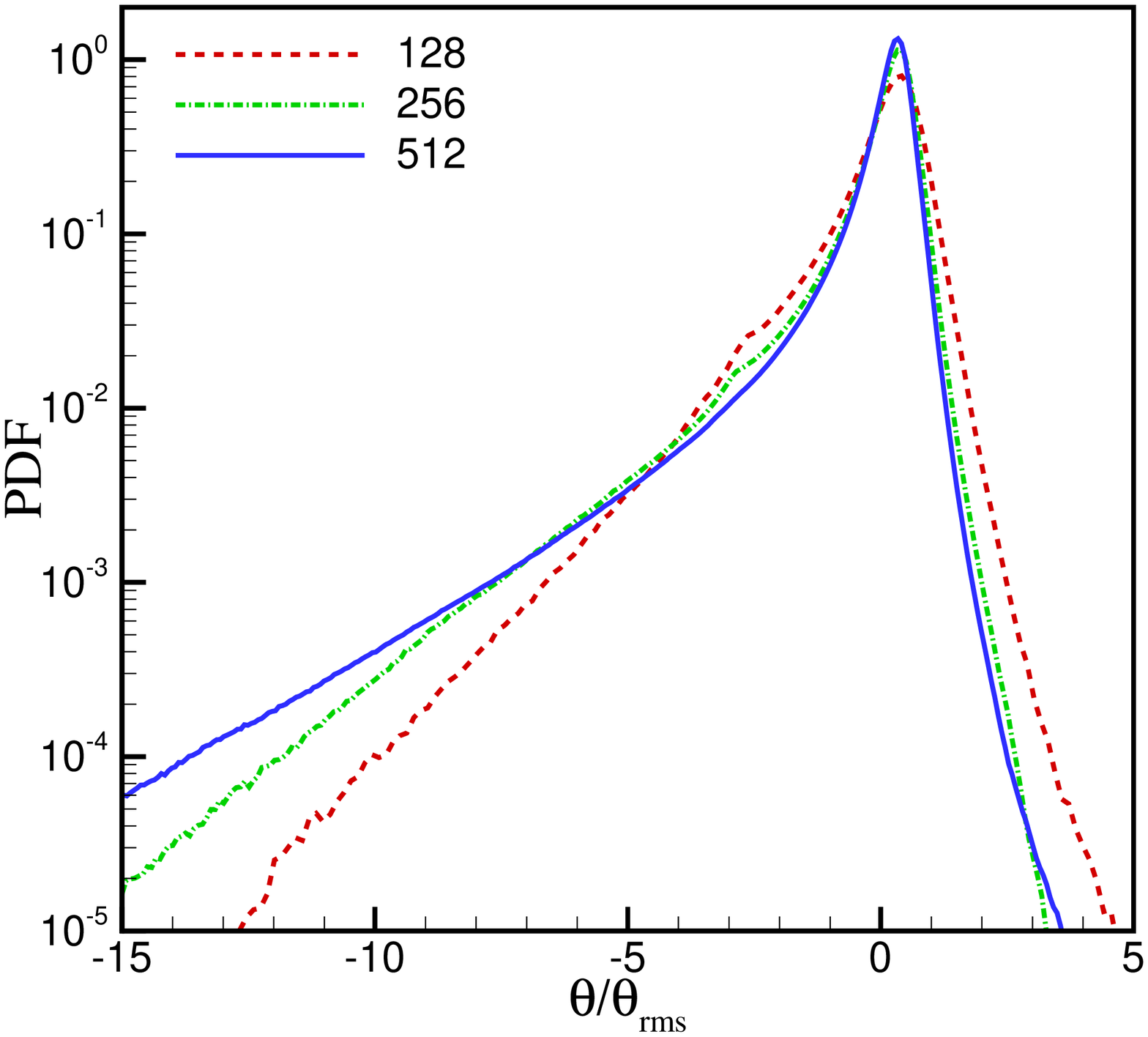}}
  \caption{Grid convergence of PDFs of normalized dilatation of stationary compressible isotropic turbulence with $R_\lambda \approx 110$, where $\theta_{\rm rms}$ denotes the root mean square of the dilatation}
  \label{fig.theta-grid}
\end{figure}

In Fig.~\ref{fig.theta-grid} we plot PDFs of normalized dilatation for different grid resolutions. The results show that the PDFs of $256^3$ and $512^3$ merge to each other very well, indicating again that the results are already convergent at the resolution of $256^3$. Note that the PDFs of the dilatation in both flows have very long negative tails, which are the direct results of shocklets, the most significant flow structures of compressible turbulence. In particular, the proportion of negative tail of $M=0.90$ is larger than that of $M=0.45$, indicating that shocklets appear in the former more frequently than in the latter (see also Fig.~\ref{fig.theta-3D} below).

\subsubsection{Preliminary comparisons of compressible isotropic turbulence}

Although the computations have already convergent at the $256^3$ resolution, we give here some preliminary comparisons of compressible turbulence simulated at the $512^3$ resolution since our present focus is to illustrate the effectiveness of Liu's scheme. More detailed analysis of compressible flow fields with turbulent Mach number $M_t$ larger than unity, namely $M_t \in [1.0, 2.6]$, but with the $256^3$ resolution, will be performed in the near future and published elsewhere.

\begin{figure}
  \centering
  \includegraphics[width=0.49\textwidth]{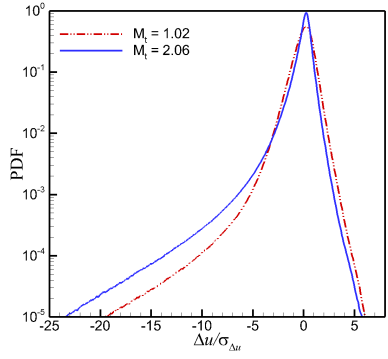}
  \caption{Averaged PDFs of longitudinal velocity increment at a separation equal to $\Delta x$ of stationary compressible isotropic turbulence with $R_\lambda = 106$, simulated with $512^3$ grid resolution, where $\sigma_{\Delta u}$ denotes the standard deviation of $\Delta u$}
  \label{fig.velM12}
\end{figure}

\begin{figure}
  \centering
  \subfigure[$M_t = 1.02$]{\includegraphics[width=0.49\textwidth]{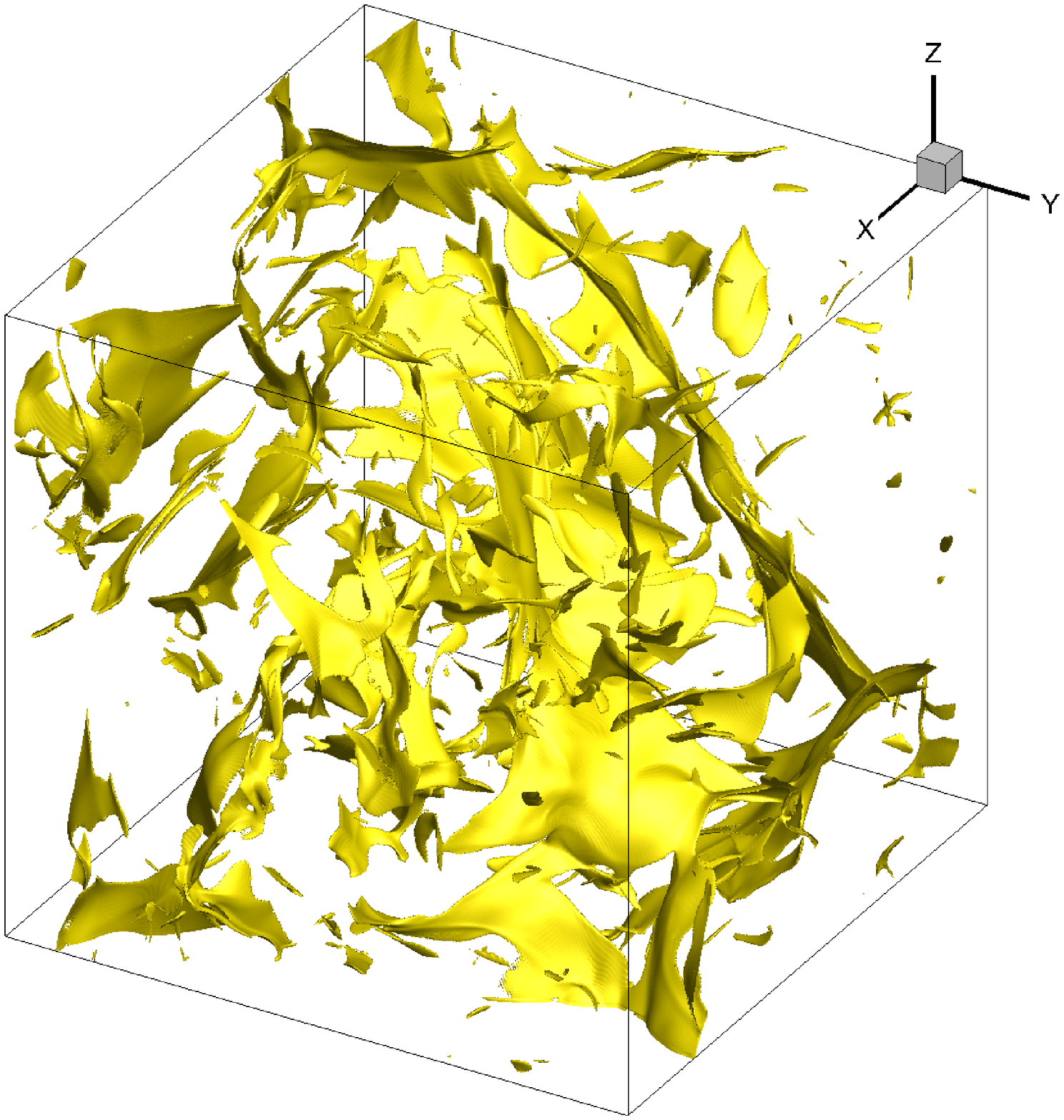}}
  \subfigure[$M_t = 2.06$]{\includegraphics[width=0.49\textwidth]{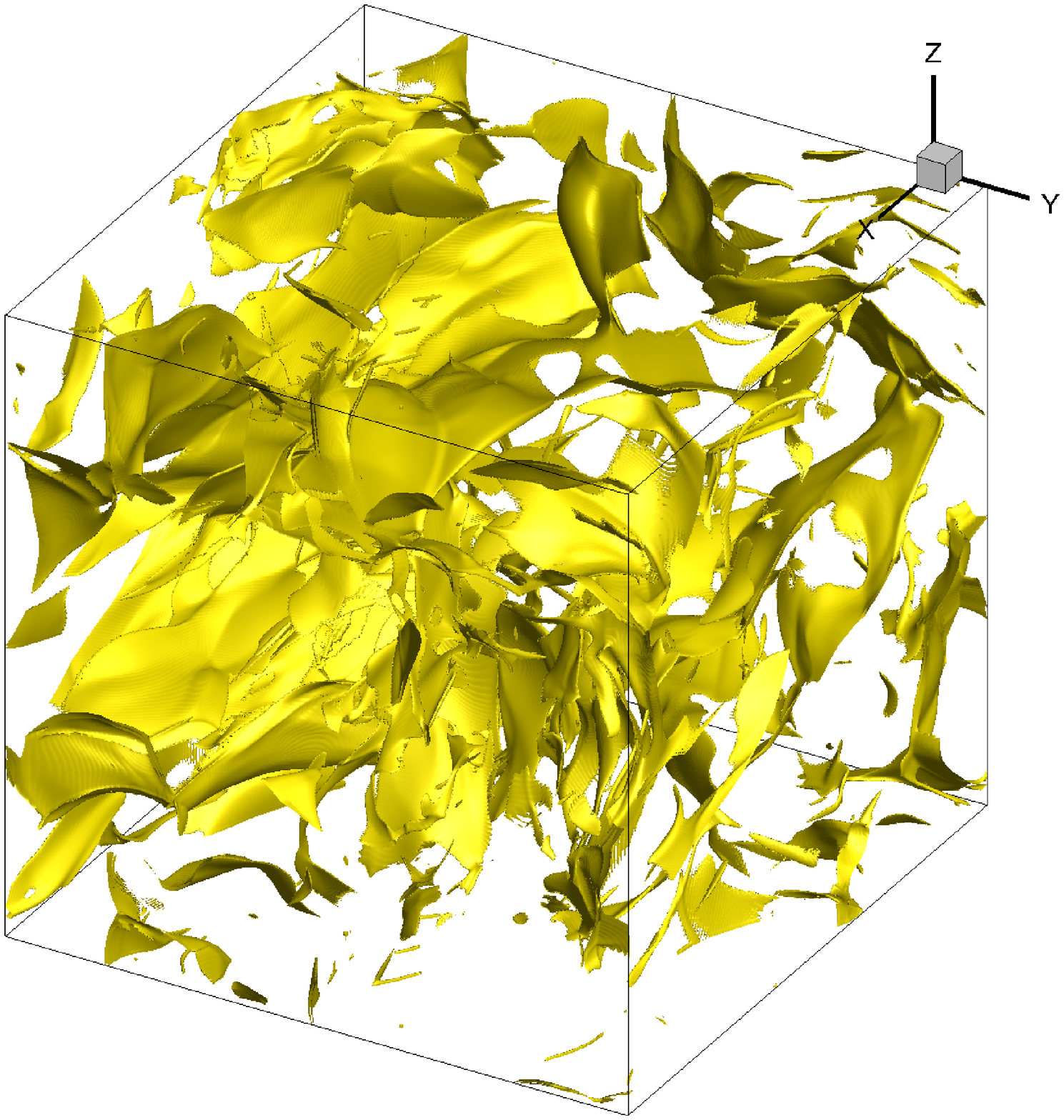}}
  \caption{Instantaneous iso-surface of dilatation of stationary compressible isotropic turbulence with $R_\lambda = 106, \theta/\theta_{\rm rms} = -3$, simulated with $512^3$ grid resolution}
  \label{fig.theta-3D}
\end{figure}

\begin{figure}
  \centering
  \subfigure[$M_t = 1.02$]{\includegraphics[width=0.49\textwidth]{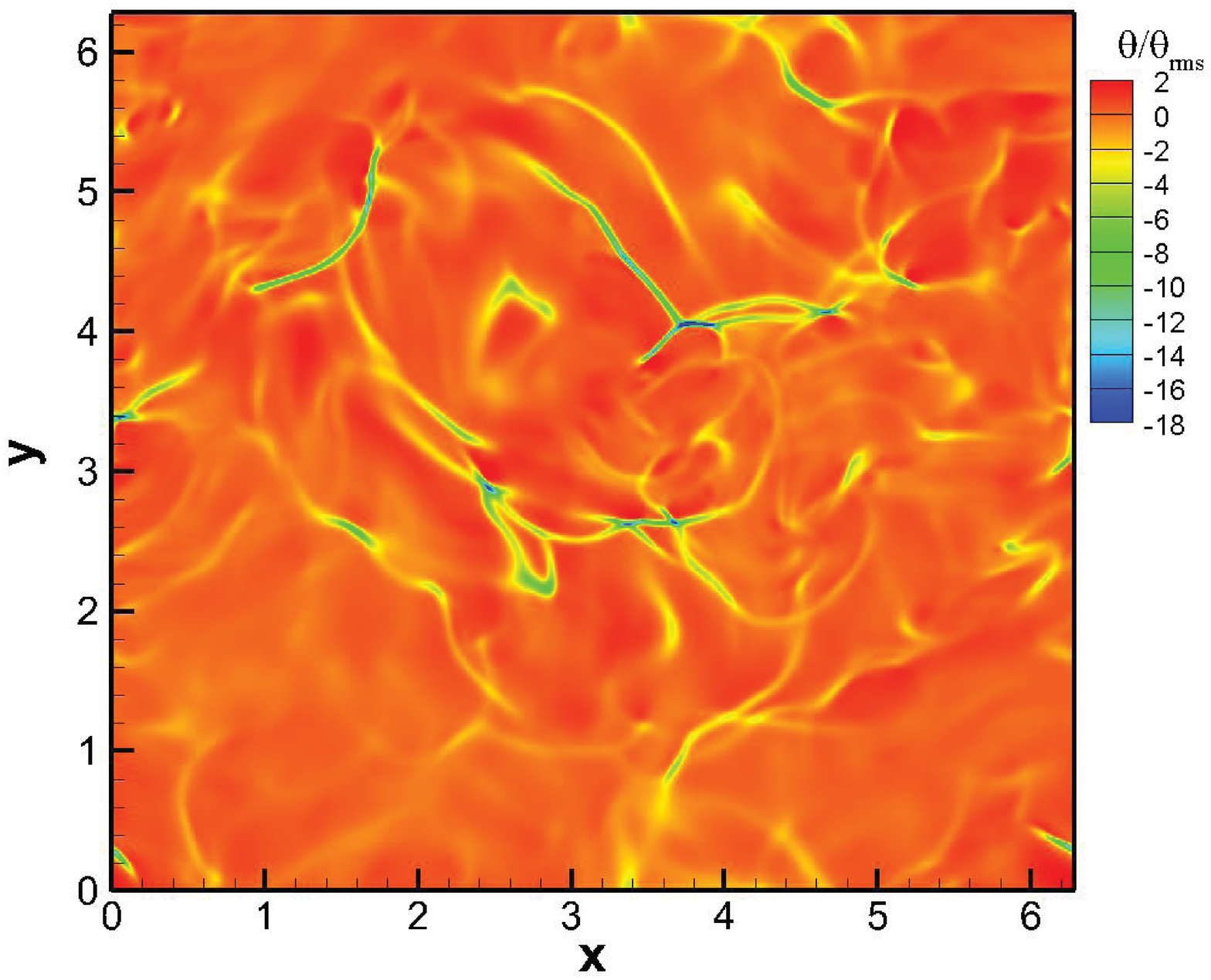}}
  \subfigure[$M_t = 2.06$]{\includegraphics[width=0.49\textwidth]{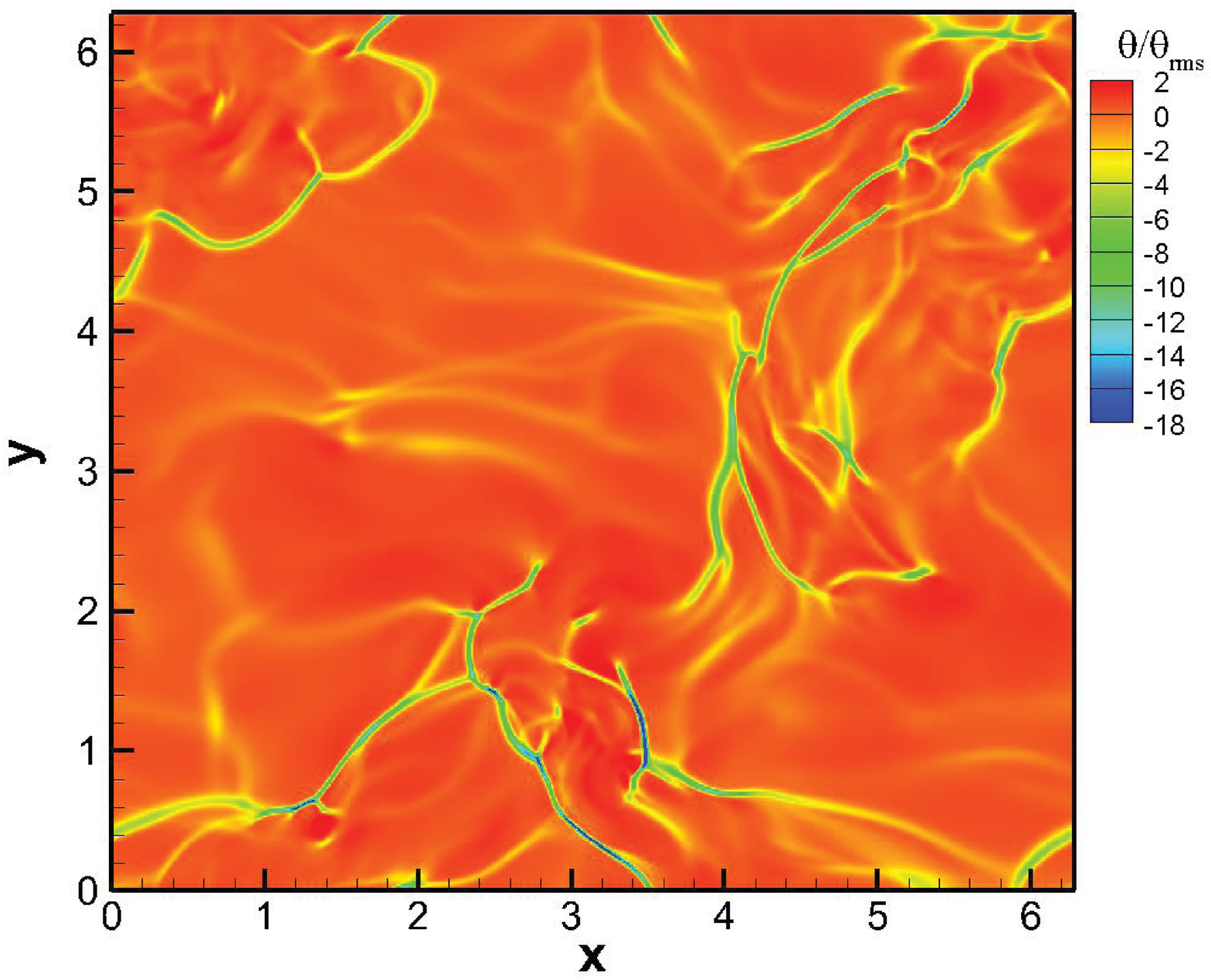}}
  \caption{Instantaneous dilatation contour on a slice $(z=\pi)$ of stationary compressible isotropic turbulence with $R_\lambda = 106$, simulated with $512^3$ grid resolution}
  \label{fig.theta-slice}
\end{figure}

Fig.~\ref{fig.velM12} shows the comparison of averaged PDFs of longitudinal velocity increment at a separation equal to $\Delta x$ with $M_t = 1.02$ and $2.06$. Evidently, the latter case with $M_t=2.06$ has a significant larger negative tail than that of the former, which implies a larger proportion of compression regions. This is due to the fact, as mentioned above, that \textit{shocklets appear more frequently in compressible turbulence with a larger turbulent Mach number}. To visualize this observation, Fig.~\ref{fig.theta-3D} shows the corresponding instantaneous iso-surface of dilatation field with $\theta/\theta_{\rm rms} = -3$. Obviously, the amount of shocklets in the whole flow field of the case $M_t=2.06$ is larger than that of $M_t=1.02$. This is even true on an arbitrary but the same slice for these two flows, which has been confirmed by Fig.~\ref{fig.theta-slice}, where the corresponding instantaneous dilatation contour on a slice is shown.

\begin{figure}
  \centering
  \includegraphics[width=0.49\textwidth]{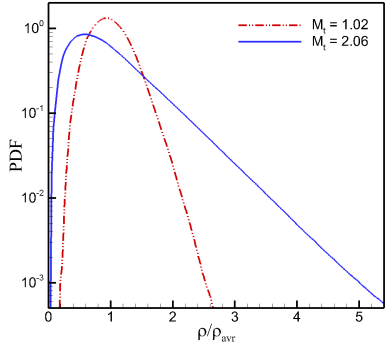}
  \caption{Averaged PDFs of local density of stationary compressible isotropic turbulence with $R_\lambda = 106$, simulated with $512^3$ grid resolution, where $\rho_{\rm avr}$ denotes the averaged density}
  \label{fig.rhoM12}
\end{figure}

A direct consequence of more shocklets is that the thermodynamic quantities have much larger fluctuations and very small density or temperature may appear. To see this conclusion more clearly, Fig.~\ref{fig.rhoM12} shows the comparison of averaged PDFs of local density for $M_t=1.02$ and $2.06$. Since the strength of shocklet increases with $M_t$ becoming larger, there is quite a lot of regions with very large density. On the other hand, what's more interesting and important, there is also a significant amount of regions where density is quite lower. For example, for the case $M_t = 2.06$ and $R_\lambda=106$ the density can be as small as 0.02 times the averaged density. Actually, it is the appearance of such large fluctuations of thermodynamic quantities that may lead to numerical simulation blow-up. Nevertheless, using the present scheme we have achieved the largest $M_t=2.6$ with the $256^3$ resolution by setting $M=1.05$ (figures not shown here). This confirms the effectiveness of the present scheme to simulate compressible turbulence with a much wider range of turbulent Mach number.

\section{Conclusions}\label{sec.5}

In this paper, the hybrid scheme proposed by \citet{Wang2010} has been improved and extended to compressible turbulence with a wider range of computable turbulent Mach number $M_t$. To achieve this goal, some techniques have been utilized. Firstly, in order to reduce nonphysical oscillations, the original hybrid scheme has been modified to the conservation form and the characteristic-wise reconstruction has been adopted. The latter is necessary to obtain a monotonic flux. Secondly, the recursive-order-reduction (ROR) method has been applied to the WENO sub-scheme, where the reconstruction-failure-detection criterion is constructed based on the idea of positivity-preserving. This sub-scheme is very effective to capture shocklets or shock waves with sufficient resolution and accuracy. Thirdly, the global Lax-Friedrichs flux has been incorporated to capture the upwind direction properly. Finally, a new cooling function has been proposed, which has been further proved to be also positivity-preserving.

To confirm that the new scheme has inherited the good characteristics of the original one and extended the computable range of turbulent Mach number, several numerical simulations have been conducted. The one-dimensional laminar problems show that the characteristic-wise reconstruction can indeed reduce numerical oscillations and the ROR-WENO scheme is applicable to extreme simulations. The three-dimensional turbulent problems illustrate that the results obtained by the new scheme and the old one are consistent with each other and what's most important is that the former can simulate flows with turbulent Mach number $M_t$ larger than unity, say, $M_t = 2.06$, of which the grid convergence has also been confirmed.

With the present scheme, one can obtain numerical data of supersonic turbulence of viscous fluid and make detailed flow analyses, which can help to deepen our understandings of compressible turbulence. More researches along this direction are under consideration and will be published elsewhere.

\section*{Acknowledgments}
\addcontentsline{toc}{section}{Acknowledgements}

This work was supported by National Natural Science Foundation of China (Grant Nos. 11702127, 11521091) and Science Challenge Project (No. TZ2016001). The numerical computation was partially performed on TianHe-1(A) at the National Supercomputer Center in Guangzhou. The authors wish to thank Prof. Yan-Tao Yang, who made a lot of insightful comments and advices on this paper. The authors would also like to thank Profs. Zuo-Li Xiao and Xin-Liang Li for their very valuable discussions. The first author gratefully acknowledges the support of Boya Postdoctoral Fellowship.

\section*{References}
\addcontentsline{toc}{section}{References}

\bibliographystyle{model1-num-names}
\bibliography{refs}

\begin{thebibliography}{71}
\expandafter\ifx\csname natexlab\endcsname\relax\def\natexlab#1{#1}\fi
\providecommand{\url}[1]{\texttt{#1}}
\providecommand{\href}[2]{#2}
\providecommand{\path}[1]{#1}
\providecommand{\DOIprefix}{doi:}
\providecommand{\ArXivprefix}{arXiv:}
\providecommand{\URLprefix}{URL: }
\providecommand{\Pubmedprefix}{pmid:}
\providecommand{\doi}[1]{\href{http://dx.doi.org/#1}{\path{#1}}}
\providecommand{\Pubmed}[1]{\href{pmid:#1}{\path{#1}}}
\providecommand{\bibinfo}[2]{#2}
\ifx\xfnm\relax \def\xfnm[#1]{\unskip,\space#1}\fi
\bibitem[{Elmegreen and Scalo(2004)}]{Elmegreen2004}
\bibinfo{author}{B.~G. Elmegreen}, \bibinfo{author}{J.~Scalo},
\newblock \bibinfo{title}{Interstellar turbulence i: Observations and
  processes},
\newblock \bibinfo{journal}{Annu. Rev. Astron. Astrophys.} \bibinfo{volume}{42}
  (\bibinfo{year}{2004}) \bibinfo{pages}{211--273}.
\bibitem[{Scalo and Elmegreen(2004)}]{Scalo2004}
\bibinfo{author}{J.~Scalo}, \bibinfo{author}{B.~G. Elmegreen},
\newblock \bibinfo{title}{Interstellar turbulence ii: Implications and
  effects},
\newblock \bibinfo{journal}{Annu. Rev. Astron. Astrophys.} \bibinfo{volume}{42}
  (\bibinfo{year}{2004}) \bibinfo{pages}{275--316}.
\bibitem[{Alexandrova et~al.(2013)Alexandrova, Chen, Sorriso-Valvo, Horbury,
  and Bale}]{Alexandrova2013}
\bibinfo{author}{O.~Alexandrova}, \bibinfo{author}{C.~H.~K. Chen},
  \bibinfo{author}{L.~Sorriso-Valvo}, \bibinfo{author}{T.~S. Horbury},
  \bibinfo{author}{S.~D. Bale},
\newblock \bibinfo{title}{Solar wind turbulence and the role of ion
  instabilities},
\newblock \bibinfo{journal}{Space Sci. Rev.} \bibinfo{volume}{178}
  (\bibinfo{year}{2013}) \bibinfo{pages}{101--139}.
\bibitem[{Mckee and Ostriker(2007)}]{Mckee2007}
\bibinfo{author}{C.~F. Mckee}, \bibinfo{author}{E.~C. Ostriker},
\newblock \bibinfo{title}{Theory of star formation},
\newblock \bibinfo{journal}{Annu. Rev. Astron. Astrophys.} \bibinfo{volume}{45}
  (\bibinfo{year}{2007}) \bibinfo{pages}{565--687}.
\bibitem[{Pope(1991)}]{Pope1991}
\bibinfo{author}{S.~B. Pope},
\newblock \bibinfo{title}{Computations of turbulent combustion: Progress and
  challenges},
\newblock \bibinfo{journal}{Symposium on Combustion} \bibinfo{volume}{23}
  (\bibinfo{year}{1991}) \bibinfo{pages}{591--612}.
\bibitem[{Ingenito and Bruno(2010)}]{Ingenito2010}
\bibinfo{author}{A.~Ingenito}, \bibinfo{author}{C.~Bruno},
\newblock \bibinfo{title}{Physics and regimes of supersonic combustion},
\newblock \bibinfo{journal}{AIAA J.} \bibinfo{volume}{48}
  (\bibinfo{year}{2010}) \bibinfo{pages}{515--525}.
\bibitem[{Lindl(1998)}]{Lindl1998}
\bibinfo{author}{J.~D. Lindl}, \bibinfo{title}{Inertial Confinement Fusion: The
  Quest for Ignition and Energy Gain Using Indirect Drive},
  \bibinfo{publisher}{Springer}, \bibinfo{address}{New York},
  \bibinfo{year}{1998}.
\bibitem[{He et~al.(2016)He, Li, Fan, Wang, Liu, Lan, Wu, and Ye}]{He2016}
\bibinfo{author}{X.~T. He}, \bibinfo{author}{J.~W. Li}, \bibinfo{author}{Z.~F.
  Fan}, \bibinfo{author}{L.~F. Wang}, \bibinfo{author}{J.~Liu},
  \bibinfo{author}{K.~Lan}, \bibinfo{author}{J.~F. Wu}, \bibinfo{author}{W.~H.
  Ye},
\newblock \bibinfo{title}{A hybrid-drive nonisobaric-ignition scheme for
  inertial confinement fusion},
\newblock \bibinfo{journal}{Phys. Plasmas} \bibinfo{volume}{23}
  (\bibinfo{year}{2016}) \bibinfo{pages}{082706}.
\bibitem[{Orszag and Patterson(1972)}]{Orszag1972}
\bibinfo{author}{S.~A. Orszag}, \bibinfo{author}{G.~S. Patterson},
\newblock \bibinfo{title}{Numerical simulation of three-dimensional homogeneous
  isotropic turbulence},
\newblock \bibinfo{journal}{Phys. Rev. Lett.} \bibinfo{volume}{28}
  (\bibinfo{year}{1972}) \bibinfo{pages}{76--79}.
\bibitem[{Hewitt and Hewitt(1979)}]{Hewitt1979}
\bibinfo{author}{E.~Hewitt}, \bibinfo{author}{R.~E. Hewitt},
\newblock \bibinfo{title}{The gibbs-wilbraham phenomenon: An episode in fourier
  analysis},
\newblock \bibinfo{journal}{Archive for History of Exact Sciences}
  \bibinfo{volume}{21} (\bibinfo{year}{1979}) \bibinfo{pages}{129--160}.
\bibitem[{Moretti(1987)}]{Moretti1987}
\bibinfo{author}{G.~Moretti},
\newblock \bibinfo{title}{Computation of flows with shocks},
\newblock \bibinfo{journal}{Annu. Rev. Fluid Mech.} \bibinfo{volume}{19}
  (\bibinfo{year}{1987}) \bibinfo{pages}{313--337}.
\bibitem[{Godunov(1959)}]{Godunov1959}
\bibinfo{author}{S.~K. Godunov},
\newblock \bibinfo{title}{A difference scheme for numerical computation of
  discontinuous solution of hydrodynamic equations},
\newblock \bibinfo{journal}{Rossiiskaya Akademiya Nauk. Matematicheskii
  Sbornik} \bibinfo{volume}{47} (\bibinfo{year}{1959})
  \bibinfo{pages}{271--306}.
\bibitem[{van Leer(1979)}]{Leer1979}
\bibinfo{author}{B.~van Leer},
\newblock \bibinfo{title}{Towards the ultimate conservative difference scheme.
  v. a second-order sequel to godunov's method},
\newblock \bibinfo{journal}{J. Comput. Phys.} \bibinfo{volume}{32}
  (\bibinfo{year}{1979}) \bibinfo{pages}{101--136}.
\bibitem[{Colella and Woodward(1984)}]{Colella1984}
\bibinfo{author}{P.~Colella}, \bibinfo{author}{P.~R. Woodward},
\newblock \bibinfo{title}{The piecewise parabolic method (ppm) for
  gas-dynamical simulations},
\newblock \bibinfo{journal}{J. Comput. Phys.} \bibinfo{volume}{54}
  (\bibinfo{year}{1984}) \bibinfo{pages}{174--201}.
\bibitem[{Toro(2009)}]{Toro2009}
\bibinfo{author}{E.~F. Toro}, \bibinfo{title}{Riemann Solvers and Numerical
  Methods for Fluid Dynamics: A Practical Introduction},
  \bibinfo{publisher}{Springer}, \bibinfo{address}{Berlin},
  \bibinfo{year}{2009}. \DOIprefix\doi{10.1007/b79761_8}.
\bibitem[{Courant et~al.(1952)Courant, Isaacson, and Rees}]{Courant1952}
\bibinfo{author}{R.~Courant}, \bibinfo{author}{E.~Isaacson},
  \bibinfo{author}{M.~Rees},
\newblock \bibinfo{title}{On the solution of nonlinear hyperbolic differential
  equations by finite differences},
\newblock \bibinfo{journal}{Comm. Pure Appl. Math.} \bibinfo{volume}{5}
  (\bibinfo{year}{1952}) \bibinfo{pages}{243--255}.
\bibitem[{Harten(1983)}]{Harten1983}
\bibinfo{author}{A.~Harten},
\newblock \bibinfo{title}{High resolution schemes for hyperbolic conservation
  laws},
\newblock \bibinfo{journal}{J. Comput. Phys.} \bibinfo{volume}{49}
  (\bibinfo{year}{1983}) \bibinfo{pages}{357--393}.
\bibitem[{Harten et~al.(1987)Harten, Engquist, Osher, and
  Chakravarthy}]{Harten1987}
\bibinfo{author}{A.~Harten}, \bibinfo{author}{B.~Engquist},
  \bibinfo{author}{S.~Osher}, \bibinfo{author}{S.~R. Chakravarthy},
\newblock \bibinfo{title}{Uniformly high order accurate essentially
  nonoscillatory schemes, iii},
\newblock \bibinfo{journal}{J. Comput. Phys.} \bibinfo{volume}{71}
  (\bibinfo{year}{1987}) \bibinfo{pages}{231--303}.
\bibitem[{Liu et~al.(1994)Liu, Osher, and Chan}]{Liu1994}
\bibinfo{author}{X.~D. Liu}, \bibinfo{author}{S.~Osher},
  \bibinfo{author}{T.~Chan},
\newblock \bibinfo{title}{Weighted essentially non-oscillatory schemes},
\newblock \bibinfo{journal}{J. Comput. Phys.} \bibinfo{volume}{115}
  (\bibinfo{year}{1994}) \bibinfo{pages}{200--212}.
\bibitem[{Von~Neumann and Richtmyer(1950)}]{Neumann1950}
\bibinfo{author}{J.~Von~Neumann}, \bibinfo{author}{R.~D. Richtmyer},
\newblock \bibinfo{title}{A method for the numerical calculation of
  hydrodynamic shocks},
\newblock \bibinfo{journal}{J. Appl. Phys.} \bibinfo{volume}{21}
  (\bibinfo{year}{1950}) \bibinfo{pages}{232--237}.
\bibitem[{Jameson et~al.(1981)Jameson, Schmidt, and Turkel}]{Jameson1981}
\bibinfo{author}{A.~Jameson}, \bibinfo{author}{W.~Schmidt},
  \bibinfo{author}{E.~L.~I. Turkel},
\newblock \bibinfo{title}{Numerical solution of the euler equations by finite
  volume methods using runge kutta time stepping schemes},
\newblock \bibinfo{journal}{AIAA Paper 81-1259}  (\bibinfo{year}{1981}).
\bibitem[{Tadmor(1989)}]{Tadmor1989}
\bibinfo{author}{E.~Tadmor},
\newblock \bibinfo{title}{Convergence of spectral methods for nonlinear
  conservation laws},
\newblock \bibinfo{journal}{SIAM J. Numer. Anal.} \bibinfo{volume}{26}
  (\bibinfo{year}{1989}) \bibinfo{pages}{30--44}.
\bibitem[{Cook and Cabot(2004)}]{Cook2004}
\bibinfo{author}{A.~W. Cook}, \bibinfo{author}{W.~H. Cabot},
\newblock \bibinfo{title}{A high-wavenumber viscosity for high-resolution
  numerical methods},
\newblock \bibinfo{journal}{J. Comput. Phys.} \bibinfo{volume}{195}
  (\bibinfo{year}{2004}) \bibinfo{pages}{594--601}.
\bibitem[{Cook and Cabot(2005)}]{Cook2005}
\bibinfo{author}{A.~W. Cook}, \bibinfo{author}{W.~H. Cabot},
\newblock \bibinfo{title}{Hyperviscosity for shock-turbulence interactions},
\newblock \bibinfo{journal}{J. Comput. Phys.} \bibinfo{volume}{203}
  (\bibinfo{year}{2005}) \bibinfo{pages}{379--385}.
\bibitem[{Harten(1978)}]{Harten1978}
\bibinfo{author}{A.~Harten},
\newblock \bibinfo{title}{The artificial compression method for computation of
  shocks and contact discontinuities. iii. self-adjusting hybrid schemes},
\newblock \bibinfo{journal}{Math. Comput.} \bibinfo{volume}{32}
  (\bibinfo{year}{1978}) \bibinfo{pages}{363--389}.
\bibitem[{Yee et~al.(1999)Yee, Sandham, and Djomehri}]{Yee1999}
\bibinfo{author}{H.~C. Yee}, \bibinfo{author}{N.~D. Sandham},
  \bibinfo{author}{M.~J. Djomehri},
\newblock \bibinfo{title}{Low-dissipative high-order shock-capturing methods
  using characteristic-based filters},
\newblock \bibinfo{journal}{J. Comput. Phys.} \bibinfo{volume}{150}
  (\bibinfo{year}{1999}) \bibinfo{pages}{199--238}.
\bibitem[{Garnier et~al.(2001)Garnier, Sagaut, and Deville}]{Garnier2001}
\bibinfo{author}{E.~Garnier}, \bibinfo{author}{P.~Sagaut},
  \bibinfo{author}{M.~Deville},
\newblock \bibinfo{title}{A class of explicit eno filters with application to
  unsteady flows},
\newblock \bibinfo{journal}{J. Comput. Phys.} \bibinfo{volume}{170}
  (\bibinfo{year}{2001}) \bibinfo{pages}{184--204}.
\bibitem[{Yee and Sj\"ogreen(2007)}]{Yee2007}
\bibinfo{author}{H.~C. Yee}, \bibinfo{author}{B.~Sj\"ogreen},
\newblock \bibinfo{title}{Development of low dissipative high order filter
  schemes for multiscale navier¨cstokes/mhd systems},
\newblock \bibinfo{journal}{J. Comput. Phys.} \bibinfo{volume}{225}
  (\bibinfo{year}{2007}) \bibinfo{pages}{910--934}.
\bibitem[{Lee et~al.(1997)Lee, Lele, and Moin}]{Lee1997}
\bibinfo{author}{S.~Lee}, \bibinfo{author}{S.~K. Lele},
  \bibinfo{author}{P.~Moin},
\newblock \bibinfo{title}{Interaction of isotropic turbulence with shock waves:
  effect of shock strength},
\newblock \bibinfo{journal}{J. Fluid Mech.} \bibinfo{volume}{340}
  (\bibinfo{year}{1997}) \bibinfo{pages}{225--247}.
\bibitem[{Adams and Shariff(1996)}]{Adams1996}
\bibinfo{author}{N.~A. Adams}, \bibinfo{author}{K.~Shariff},
\newblock \bibinfo{title}{A high-resolution hybrid compact-eno scheme for
  shock-turbulence interaction problems},
\newblock \bibinfo{journal}{J. Comput. Phys.} \bibinfo{volume}{127}
  (\bibinfo{year}{1996}) \bibinfo{pages}{27--51}.
\bibitem[{Pirozzoli(2002)}]{Pirozzoli2002}
\bibinfo{author}{S.~Pirozzoli},
\newblock \bibinfo{title}{Conservative hybrid compact-weno schemes for
  shock-turbulence interaction},
\newblock \bibinfo{journal}{J. Comput. Phys.} \bibinfo{volume}{178}
  (\bibinfo{year}{2002}) \bibinfo{pages}{81--117}.
\bibitem[{Ren et~al.(2003)Ren, Liu, and Zhang}]{Ren2003}
\bibinfo{author}{Y.~X. Ren}, \bibinfo{author}{M.~E. Liu},
  \bibinfo{author}{H.~X. Zhang},
\newblock \bibinfo{title}{A characteristic-wise hybrid compact-weno scheme for
  solving hyperbolic conservation laws},
\newblock \bibinfo{journal}{J. Comput. Phys.} \bibinfo{volume}{192}
  (\bibinfo{year}{2003}) \bibinfo{pages}{365--386}.
\bibitem[{Wang et~al.(2010)Wang, Wang, Xiao, Shi, and Chen}]{Wang2010}
\bibinfo{author}{J.~Wang}, \bibinfo{author}{L.~P. Wang},
  \bibinfo{author}{Z.~Xiao}, \bibinfo{author}{Y.~Shi},
  \bibinfo{author}{S.~Chen},
\newblock \bibinfo{title}{A hybrid numerical simulation of isotropic
  compressible turbulence},
\newblock \bibinfo{journal}{J. Comput. Phys.} \bibinfo{volume}{229}
  (\bibinfo{year}{2010}) \bibinfo{pages}{5257--5279}.
\bibitem[{Pirozzoli(2011)}]{Pirozzoli2011}
\bibinfo{author}{S.~Pirozzoli},
\newblock \bibinfo{title}{Numerical methods for high-speed flows},
\newblock \bibinfo{journal}{Annu. Rev. Fluid Mech.} \bibinfo{volume}{43}
  (\bibinfo{year}{2011}) \bibinfo{pages}{163--194}.
\bibitem[{Johnsen et~al.(2010)Johnsen, Larsson, Bhagatwala, Cabot, Moin, Olson,
  Rawat, Shankar, Sj?green, Yee, Zhong, and L.}]{Johnsen2010}
\bibinfo{author}{E.~Johnsen}, \bibinfo{author}{J.~Larsson},
  \bibinfo{author}{A.~V. Bhagatwala}, \bibinfo{author}{W.~H. Cabot},
  \bibinfo{author}{P.~Moin}, \bibinfo{author}{B.~J. Olson},
  \bibinfo{author}{P.~S. Rawat}, \bibinfo{author}{S.~K. Shankar},
  \bibinfo{author}{B.~Sj?green}, \bibinfo{author}{H.~C. Yee},
  \bibinfo{author}{X.~L. Zhong}, \bibinfo{author}{S.~K. L.},
\newblock \bibinfo{title}{Assessment of high-resolution methods for numerical
  simulations of compressible turbulence with shock waves},
\newblock \bibinfo{journal}{J. Comput. Phys.} \bibinfo{volume}{229}
  (\bibinfo{year}{2010}) \bibinfo{pages}{1213--1237}.
\bibitem[{Zhou et~al.(2007)Zhou, Yao, He, and Shen}]{Zhou2007}
\bibinfo{author}{Q.~Zhou}, \bibinfo{author}{Z.~H. Yao},
  \bibinfo{author}{F.~He}, \bibinfo{author}{M.~Y. Shen},
\newblock \bibinfo{title}{A new family of high-order compact upwind difference
  schemes with good spectral resolution},
\newblock \bibinfo{journal}{J. Comput. Phys.} \bibinfo{volume}{227}
  (\bibinfo{year}{2007}) \bibinfo{pages}{1306--1339}.
\bibitem[{Hill and Pullin(2004)}]{Hill2004}
\bibinfo{author}{D.~J. Hill}, \bibinfo{author}{D.~I. Pullin},
\newblock \bibinfo{title}{Hybrid tuned center-difference-weno method for large
  eddy simulations in the presence of strong shocks},
\newblock \bibinfo{journal}{J. Comput. Phys.} \bibinfo{volume}{194}
  (\bibinfo{year}{2004}) \bibinfo{pages}{435--450}.
\bibitem[{Kim and Kwon(2005)}]{Kim2005}
\bibinfo{author}{D.~Kim}, \bibinfo{author}{J.~H. Kwon},
\newblock \bibinfo{title}{A high-order accurate hybrid scheme using a central
  flux scheme and a weno scheme for compressible flowfield analysis},
\newblock \bibinfo{journal}{J. Comput. Phys.} \bibinfo{volume}{210}
  (\bibinfo{year}{2005}) \bibinfo{pages}{554--583}.
\bibitem[{Larsson et~al.(2007)Larsson, Lele, and Moin}]{Larsson2007}
\bibinfo{author}{J.~Larsson}, \bibinfo{author}{S.~K. Lele},
  \bibinfo{author}{P.~Moin},
\newblock \bibinfo{title}{Effect of numerical dissipation on the predicted
  spectra for compressible turbulence},
\newblock \bibinfo{journal}{Center for Turbulence Research}
  (\bibinfo{year}{2007}).
\bibitem[{Samtaney et~al.(2001)Samtaney, Pullin, and Kosovi\'c}]{Samtaney2001}
\bibinfo{author}{R.~Samtaney}, \bibinfo{author}{D.~I. Pullin},
  \bibinfo{author}{B.~Kosovi\'c},
\newblock \bibinfo{title}{Direct numerical simulation of decaying compressible
  turbulence and shocklet statistics},
\newblock \bibinfo{journal}{Phys. Fluids} \bibinfo{volume}{13}
  (\bibinfo{year}{2001}) \bibinfo{pages}{1415--1430}.
\bibitem[{Wang et~al.(2011)Wang, Shi, Wang, Xiao, He, and Chen}]{Wang2011}
\bibinfo{author}{J.~C. Wang}, \bibinfo{author}{Y.~P. Shi},
  \bibinfo{author}{L.~P. Wang}, \bibinfo{author}{Z.~L. Xiao},
  \bibinfo{author}{X.~T. He}, \bibinfo{author}{S.~Y. Chen},
\newblock \bibinfo{title}{Effect of shocklets on the velocity gradients in
  highly compressible isotropic turbulence},
\newblock \bibinfo{journal}{Phys. Fluids} \bibinfo{volume}{23}
  (\bibinfo{year}{2011}).
\bibitem[{Wang et~al.(2012{\natexlab{a}})Wang, Shi, Wang, Xiao, He, and
  Chen}]{Wang2012-jfm}
\bibinfo{author}{J.~C. Wang}, \bibinfo{author}{Y.~P. Shi},
  \bibinfo{author}{L.~P. Wang}, \bibinfo{author}{Z.~L. Xiao},
  \bibinfo{author}{X.~T. He}, \bibinfo{author}{S.~Y. Chen},
\newblock \bibinfo{title}{Effect of compressibility on the small-scale
  structures in isotropic turbulence},
\newblock \bibinfo{journal}{J. Fluid Mech.} \bibinfo{volume}{713}
  (\bibinfo{year}{2012}{\natexlab{a}}) \bibinfo{pages}{588--631}.
\bibitem[{Wang et~al.(2012{\natexlab{b}})Wang, Shi, Wang, Xiao, He, and
  Chen}]{Wang2012-prl}
\bibinfo{author}{J.~C. Wang}, \bibinfo{author}{Y.~P. Shi},
  \bibinfo{author}{L.~P. Wang}, \bibinfo{author}{Z.~L. Xiao},
  \bibinfo{author}{X.~T. He}, \bibinfo{author}{S.~Y. Chen},
\newblock \bibinfo{title}{Scaling and statistics in three-dimensional
  compressible turbulence},
\newblock \bibinfo{journal}{Phys. Rev. Lett.} \bibinfo{volume}{108}
  (\bibinfo{year}{2012}{\natexlab{b}}) \bibinfo{pages}{214505}.
\bibitem[{Yang et~al.(2014)Yang, Wang, Shi, Xiao, He, and Chen}]{Yang2014}
\bibinfo{author}{Y.~T. Yang}, \bibinfo{author}{J.~C. Wang},
  \bibinfo{author}{Y.~P. Shi}, \bibinfo{author}{Z.~L. Xiao},
  \bibinfo{author}{X.~T. He}, \bibinfo{author}{S.~Y. Chen},
\newblock \bibinfo{title}{Interactions between inertial particles and shocklets
  in compressible turbulent flow},
\newblock \bibinfo{journal}{Phys. Fluids} \bibinfo{volume}{26}
  (\bibinfo{year}{2014}) \bibinfo{pages}{091702}.
\bibitem[{Chen et~al.(2015)Chen, Xia, Wang, and Yang}]{Chen2015}
\bibinfo{author}{S.~Y. Chen}, \bibinfo{author}{Z.~H. Xia},
  \bibinfo{author}{J.~C. Wang}, \bibinfo{author}{Y.~T. Yang},
\newblock \bibinfo{title}{Recent progress in compressible turbulence},
\newblock \bibinfo{journal}{Acta Mech. Sin.} \bibinfo{volume}{31}
  (\bibinfo{year}{2015}) \bibinfo{pages}{275--291}.
\bibitem[{Federrath(2013)}]{Federrath2013}
\bibinfo{author}{C.~Federrath},
\newblock \bibinfo{title}{On the universality of supersonic turbulence},
\newblock \bibinfo{journal}{Monthly Notices of the Royal Astronomical Society}
  \bibinfo{volume}{436} (\bibinfo{year}{2013}) \bibinfo{pages}{1245--1257}.
\bibitem[{Jagannathan and Donzis(2016)}]{Jagannathan2016}
\bibinfo{author}{S.~Jagannathan}, \bibinfo{author}{D.~A. Donzis},
\newblock \bibinfo{title}{Reynolds and mach number scaling in
  solenoidally-forced compressible turbulence using high-resolution direct
  numerical?simulations},
\newblock \bibinfo{journal}{J. Fluid Mech.} \bibinfo{volume}{789}
  (\bibinfo{year}{2016}) \bibinfo{pages}{669--707}.
\bibitem[{Pirozzoli and Grasso(2004)}]{Pirozzoli2004}
\bibinfo{author}{S.~Pirozzoli}, \bibinfo{author}{F.~Grasso},
\newblock \bibinfo{title}{Direct numerical simulations of isotropic
  compressible turbulence: Influence of compressibility on dynamics and
  structures},
\newblock \bibinfo{journal}{Phys. Fluids} \bibinfo{volume}{16}
  (\bibinfo{year}{2004}) \bibinfo{pages}{4386--4407}.
\bibitem[{Lee et~al.(2009)Lee, Girimaji, and Kerimo}]{Lee2009}
\bibinfo{author}{K.~Lee}, \bibinfo{author}{S.~S. Girimaji},
  \bibinfo{author}{J.~Kerimo},
\newblock \bibinfo{title}{Effect of compressibility on turbulent velocity
  gradients and small-scale structure},
\newblock \bibinfo{journal}{J. Turbul.} \bibinfo{volume}{10}
  (\bibinfo{year}{2009}) \bibinfo{pages}{1--18}.
\bibitem[{Xia et~al.(2016)Xia, Shi, Zhang, and Chen}]{Xia2016}
\bibinfo{author}{Z.~H. Xia}, \bibinfo{author}{Y.~P. Shi},
  \bibinfo{author}{Q.~Q. Zhang}, \bibinfo{author}{S.~Y. Chen},
\newblock \bibinfo{title}{Modulation to compressible homogenous turbulence by
  heavy point particles. i. effect of particles¡¯ density},
\newblock \bibinfo{journal}{Phys. Fluids} \bibinfo{volume}{28}
  (\bibinfo{year}{2016}) \bibinfo{pages}{016103}.
\bibitem[{Sutherland(1893)}]{Sutherland1893}
\bibinfo{author}{D.~M. Sutherland},
\newblock \bibinfo{title}{The viscosity of gases and molecular force},
\newblock \bibinfo{journal}{Philos. Mag.} \bibinfo{volume}{5}
  (\bibinfo{year}{1893}) \bibinfo{pages}{507--531}.
\bibitem[{Lunev(2009)}]{Lunev2009}
\bibinfo{author}{V.~V. Lunev}, \bibinfo{title}{Real Gas Flows with High
  Velocities}, \bibinfo{publisher}{CRC}, \bibinfo{address}{New York},
  \bibinfo{year}{2009}.
\bibitem[{Chen et~al.(1993)Chen, Doolen, Kraichnan, and She}]{Chen1993}
\bibinfo{author}{S.~Y. Chen}, \bibinfo{author}{G.~D. Doolen},
  \bibinfo{author}{R.~H. Kraichnan}, \bibinfo{author}{Z.~S. She},
\newblock \bibinfo{title}{On statistical correlations between velocity
  increments and locally averaged dissipation in homogeneous turbulence},
\newblock \bibinfo{journal}{Phys. Fluids} \bibinfo{volume}{5}
  (\bibinfo{year}{1993}) \bibinfo{pages}{458}.
\bibitem[{Passot et~al.(1995)Passot, Vazquezsemadeni, and Pouquet}]{Passot1995}
\bibinfo{author}{T.~Passot}, \bibinfo{author}{E.~Vazquezsemadeni},
  \bibinfo{author}{A.~Pouquet},
\newblock \bibinfo{title}{A turbulent model for the interstellar medium. ii.
  magnetic fields and rotation},
\newblock \bibinfo{journal}{Astrophysical Journal} \bibinfo{volume}{455}
  (\bibinfo{year}{1995}) \bibinfo{pages}{536--555}.
\bibitem[{Balsara and Shu(2000)}]{Balsara2000}
\bibinfo{author}{D.~S. Balsara}, \bibinfo{author}{C.~W. Shu},
\newblock \bibinfo{title}{Monotonicity preserving weighted essentially
  non-oscillatory schemes with increasingly high order of accuracy},
\newblock \bibinfo{journal}{J. Comput. Phys.} \bibinfo{volume}{160}
  (\bibinfo{year}{2000}) \bibinfo{pages}{405--452}.
\bibitem[{Lele(1992)}]{Lele1992}
\bibinfo{author}{S.~K. Lele},
\newblock \bibinfo{title}{Compact finite-difference schemes with spectral-like
  resolution},
\newblock \bibinfo{journal}{J. Comput. Phys.} \bibinfo{volume}{103}
  (\bibinfo{year}{1992}) \bibinfo{pages}{16--42}.
\bibitem[{Steger and Warming(1981)}]{Steger1981}
\bibinfo{author}{J.~L. Steger}, \bibinfo{author}{R.~F. Warming},
\newblock \bibinfo{title}{Flux vector splitting of inviscid gasdynamic
  equations with application to finite difference methods},
\newblock \bibinfo{journal}{J. Comput. Phys.} \bibinfo{volume}{40}
  (\bibinfo{year}{1981}) \bibinfo{pages}{263--293}.
\bibitem[{Shu and Osher(1988)}]{Shu1988}
\bibinfo{author}{C.~W. Shu}, \bibinfo{author}{S.~Osher},
\newblock \bibinfo{title}{Efficient implementation of essentially
  non-oscillatory shock-capturing schemes},
\newblock \bibinfo{journal}{J. Comput. Phys.} \bibinfo{volume}{77}
  (\bibinfo{year}{1988}) \bibinfo{pages}{439--471}.
\bibitem[{Phillips(1959)}]{Phillips1959}
\bibinfo{author}{N.~A. Phillips}, \bibinfo{title}{An example of non-linear
  computational instability}, \bibinfo{publisher}{Oxford Univ. Press},
  \bibinfo{address}{Oxford}, \bibinfo{year}{1959}.
\bibitem[{Lax(1954)}]{Lax1954}
\bibinfo{author}{P.~D. Lax},
\newblock \bibinfo{title}{Weak solutions of nonlinear hyperbolic equations and
  their numerical computation},
\newblock \bibinfo{journal}{Comm. Pure Appl. Math.} \bibinfo{volume}{7}
  (\bibinfo{year}{1954}) \bibinfo{pages}{159--193}.
\bibitem[{Harten et~al.(1986)Harten, Osher, Engquist, and
  Chakravarthy}]{Harten1986}
\bibinfo{author}{A.~Harten}, \bibinfo{author}{S.~Osher},
  \bibinfo{author}{B.~Engquist}, \bibinfo{author}{S.~R. Chakravarthy},
\newblock \bibinfo{title}{Some results on uniformly high-order accurate
  essentially nonoscillatory schemes},
\newblock \bibinfo{journal}{Appl. Numer. Math.} \bibinfo{volume}{2}
  (\bibinfo{year}{1986}) \bibinfo{pages}{347--377}.
\bibitem[{Titarev and Toro(2004)}]{Titarev2004}
\bibinfo{author}{V.~A. Titarev}, \bibinfo{author}{E.~F. Toro},
\newblock \bibinfo{title}{Finite-volume weno schemes for three-dimensional
  conservation laws},
\newblock \bibinfo{journal}{J. Comput. Phys.} \bibinfo{volume}{201}
  (\bibinfo{year}{2004}) \bibinfo{pages}{238--260}.
\bibitem[{Gerolymos et~al.(2009)Gerolymos, S¨¦n¨¦chal, and
  Vallet}]{Gerolymos2009}
\bibinfo{author}{G.~A. Gerolymos}, \bibinfo{author}{D.~S¨¦n¨¦chal},
  \bibinfo{author}{I.~Vallet},
\newblock \bibinfo{title}{Very-high-order weno schemes},
\newblock \bibinfo{journal}{J. Comput. Phys.} \bibinfo{volume}{228}
  (\bibinfo{year}{2009}) \bibinfo{pages}{8481--8524}.
\bibitem[{Shu(2016)}]{Shu2016}
\bibinfo{author}{C.~W. Shu},
\newblock \bibinfo{title}{High order weno and dg methods for time-dependent
  convection-dominated pdes: A brief survey of several recent developments},
\newblock \bibinfo{journal}{J. Comput. Phys.} \bibinfo{volume}{316}
  (\bibinfo{year}{2016}) \bibinfo{pages}{598--613}.
\bibitem[{Hu et~al.(2013)Hu, Adams, and Shu}]{Hu2013}
\bibinfo{author}{X.~Y. Hu}, \bibinfo{author}{N.~A. Adams},
  \bibinfo{author}{C.~W. Shu},
\newblock \bibinfo{title}{Positivity-preserving method for high-order
  conservative schemes solving compressible euler equations},
\newblock \bibinfo{journal}{J. Comput. Phys.} \bibinfo{volume}{242}
  (\bibinfo{year}{2013}) \bibinfo{pages}{169--180}.
\bibitem[{Roe(1981)}]{Roe1981}
\bibinfo{author}{P.~L. Roe},
\newblock \bibinfo{title}{Approximate riemann solvers, parameter vectors, and
  difference schemes},
\newblock \bibinfo{journal}{J. Comput. Phys.} \bibinfo{volume}{43}
  (\bibinfo{year}{1981}) \bibinfo{pages}{357--372}.
\bibitem[{Osher(1981)}]{Osher1981}
\bibinfo{author}{S.~Osher}, \bibinfo{title}{Numerical Solution of Singular
  Perturbation Problems and Hyperbolic Systems of Conservation Laws},
  volume~\bibinfo{volume}{47}, \bibinfo{publisher}{North-Holland},
  \bibinfo{year}{1981}, pp. \bibinfo{pages}{179--204}.
  \DOIprefix\doi{10.1016/S0304-0208(08)71109-5}.
\bibitem[{van Leer(1982)}]{Leer1982}
\bibinfo{author}{B.~van Leer}, \bibinfo{title}{Flux-vector splitting for the
  Euler equations}, \bibinfo{publisher}{Springer}, \bibinfo{address}{Berlin},
  \bibinfo{year}{1982}, pp. \bibinfo{pages}{507--512}.
  \DOIprefix\doi{10.1007/3-540-11948-5_66}.
\bibitem[{Sanders and Prendergast(1974)}]{Sanders1974}
\bibinfo{author}{R.~H. Sanders}, \bibinfo{author}{K.~H. Prendergast},
\newblock \bibinfo{title}{The possible relation of the 3-kiloparsec arm to
  explosions in the galactic nucleus},
\newblock \bibinfo{journal}{Astrophysical Journal} \bibinfo{volume}{188}
  (\bibinfo{year}{1974}) \bibinfo{pages}{489--500}.
\bibitem[{Sod(1978)}]{Sod1978}
\bibinfo{author}{G.~A. Sod},
\newblock \bibinfo{title}{A survey of several finite difference methods for
  systems of nonlinear hyperbolic conservation laws},
\newblock \bibinfo{journal}{J. Comput. Phys.} \bibinfo{volume}{27}
  (\bibinfo{year}{1978}) \bibinfo{pages}{1--31}.
\bibitem[{Woodward and Colella(1984)}]{Woodward1984}
\bibinfo{author}{P.~Woodward}, \bibinfo{author}{P.~Colella},
\newblock \bibinfo{title}{The numerical simulation of two-dimensional fluid
  flow with strong shocks},
\newblock \bibinfo{journal}{J. Comput. Phys.} \bibinfo{volume}{54}
  (\bibinfo{year}{1984}) \bibinfo{pages}{115--173}.

\end{thebibliography}

\end{document}